\documentclass[aps,pre,floats,superscriptaddress,showpacs,twocolumn]{revtex4}
\usepackage{amsmath, amsthm, amssymb}
\usepackage{enumitem}
\usepackage{graphics,graphicx}
\usepackage{epsfig}
\usepackage{times}
\usepackage{epstopdf}
\usepackage{lipsum}
\usepackage{dcolumn}
\usepackage{bm}

\usepackage{ulem, color}

\begin{document}

\title{A framework for epidemic spreading in multiplex networks of metapopulations}

\author{D. Soriano-Pa\~nos}
\affiliation{Departamento de F\'{\i}sica de la Materia Condensada, Universidad de Zaragoza, 50009 Zaragoza, Spain}
\affiliation{GOTHAM Lab, Instituto de Biocomputaci\'on y F\'{\i}sica de Sistemas Complejos (BIFI), Universidad de Zaragoza, 50018 Zaragoza, Spain}
\author{L. Lotero}
\affiliation{Facultad de Ingenier\'{\i}a Industrial, Universidad Pontificia Bolivariana, Medell\'{\i}n, Colombia}
\author{A. Arenas}\email{alexandre.arenas@urv.cat}
\affiliation{Departament d'Enginyeria Inform\`atica i Matem\`atiques, Universitat Rovira i Virgili, 43007 Tarragona, Spain}
\author{J. G\'omez-Garde\~nes}\email{gardenes@unizar.es}
\affiliation{Departamento de F\'{\i}sica de la Materia Condensada, Universidad de Zaragoza, 50009 Zaragoza, Spain}
\affiliation{GOTHAM Lab, Instituto de Biocomputaci\'on y F\'{\i}sica de Sistemas Complejos (BIFI), Universidad de Zaragoza, 50018 Zaragoza, Spain}

\date{\today}
             
\begin{abstract}
We propose a theoretical framework for the  study of epidemics in structured metapopulations, with heterogeneous agents, subjected to recurrent mobility patterns. We propose to represent the heterogeneity in the composition of the metapopulations as layers in a multiplex network, where nodes would correspond to geographical areas and layers account for the mobility patterns of agents of the same class.
We analyze both the classical Susceptible-Infected-Susceptible and the Susceptible-Infected-Removed epidemic models within this framework, and compare macroscopic and microscopic indicators of the spreading process with extensive Monte Carlo simulations. Our results are in excellent agreement with the simulations. We also derive an exact expression of the epidemic threshold on this general framework revealing a non-trivial dependence on the mobility parameter. 
\end{abstract}

\maketitle

\section{Introduction}

During the last decades we have witnessed the onset of several major global health threats such as the 2003 spread of SARS, the H1N1 influenza pandemic in 2009, the  western Africa 2014 Ebola outbreaks and more recently the Zika epidemics in the Americas and Caribbean regions. These outbreaks are increasingly characterized by the small elapsed time between initial infections in a single region to the global epidemic state affecting different cities, regions, countries and, in some cases, continents. Thus, in the recent years a great effort has been devoted to understand the fast unfold of emergent diseases and to design both local and global contention strategies. The most common avenue to tackle this problem is to adapt classical epidemic models taking into account the multiscale nature of diseases propagation \cite{gleam1,gleam2}.

It is clear that the spread of an emergent infectious disease is the result of human-human interactions in small geographical patches. However, in order to understand the geographical diffusion of diseases, one has to combine these microscopic contagion processes with the long-range disease propagation due to human mobility across different spatial scales. To tackle this problem, epidemic modeling has relied on reaction-diffusion dynamics in metapopulations, a family of models first used in the field of population ecology \cite{hanski,metapop1,metapop2,metapop3,metapop4}. For the case of epidemic modeling, the usual metapopulation scenario \cite{metapop5,pion1,pion2} is as follows. A population is distributed in a set of patches, being the size (number of individuals) of each patch in principle different. The individuals within each patch are well-mixed, {\it i.e.}, pathogens can be transmitted from an infected host to any of the healthy agents placed in the same patch with the same probability. 
The second ingredient of metapopulation frameworks concerns the mobility of agents. Each host is allowed to change its current location and occupy another patch, thus fostering the spread of pathogens at the system level. Mobility of agents between different patches is usually represented in terms of a network where nodes are locations while a link between two patches represents the possibility of moving between them. 

The non-trivial mobility patterns observed in real populations \cite{airline} and the recent advances of network epidemiology \cite{review} have motivated a thorough analysis about the impact that the structure of mobility networks has on the onset of global-scale contagions. In the last decade, important steps towards the inclusion of realistic mobility structures have been done \cite{colizza1,colizza2,colizza3,colizza4}. These approaches had to compromise between realism and analytical feasibility. 
On one side, lengthy mechanistic simulations\cite{gleam1,eubank} provide fair predictions on realistic scenarios while, on the other, theoretical frameworks allowing for analytical results usually rely on  strong assumptions limiting their applicability to real-world threats. For instance, it is usual to assume simplified mobility patterns and mean-field approximations for hosts and patches behavior to be able to predict the onset of an outbreak. 
In this class of models, random diffusion of agents between the nodes is often used as proxy of human mobility while, as in the case of the heterogeneous mean field approach in scale-free networks \cite{hmf}, subpopulations with identical connectivity patterns are assumed to be equally affected by the disease. 

These mean-field like approximations for patches having identical properties, while useful for deriving analytical results, add important limitations for their applicability in real-world diseases prediction.  As data gathering techniques and epidemic surveillance \cite{influenzanet} increase their accuracy, metapopulation models face new challenges \cite{ball}. In an effort of relaxing the assumption about random diffusion of hosts and approaching realistic mobility patterns used in agent based simulations, recently the recurrent and spatially constrained nature of most human movements, such as daily commutes, has been addressed \cite{commutes,prx,jtb,epjb}, at the expense of considering either mean-field assumptions or simple mobility networks.  Therefore, a theoretical framework of general metapopulations of arbitrary structure, able at incorporating real mobility patterns, remains as an open theoretical challenge.

\begin{figure*}[t!]
\begin{center}
\includegraphics[width=1.37\columnwidth,angle=-0]{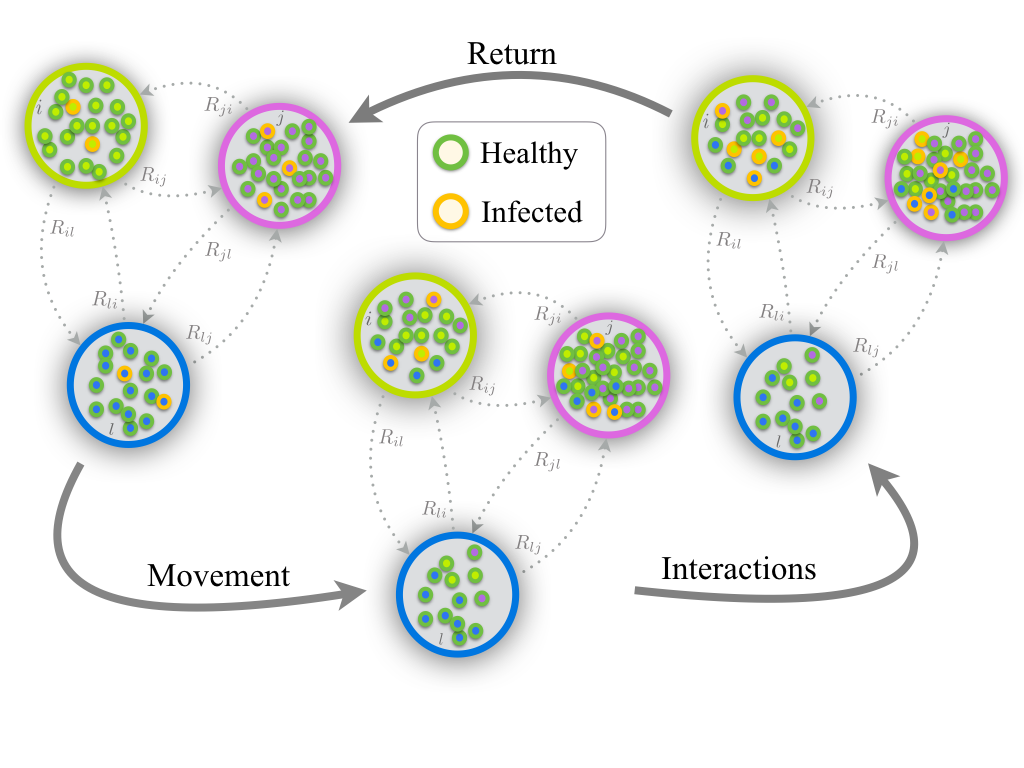}
\caption{(color online). Schematic representation of one time step of the {\it Movement}-{\it Interaction}-{\it Return} (MIR) metapopulation model. The network is composed of $N=3$ patches. 
At the {\it movement} stage some of the local agents decide to move to the other patches according to the probabilities encoded in matrix ${\bf R}$. Once agents have moved interact in a well-mixed way and change their epidemic status (Healthy or Infected) according to an SIS model. Finally, the agents come back to their home patches and a new time step starts.}
\label{fig:MIR}
\end{center}
\end{figure*}

In addition, very recently, network science has tackled the formulation of networked systems in which different types of interactions between a given set of nodes coexist and interplay. These systems, termed as multiplex networks \cite{kivela2014,boccaletti2014,dedomenico2013,battiston,radicchi}, consist on a set of $L$ networks (usually called layers) and a set of $N$ nodes. Each node is represented once in each network layer allowing it to share different connectivity patterns in each of the $L$ layers. In terms of metapopulation models, for which nodes account for geographical locations, the multiplex formalism captures the coexistence, within each subpopulation, of different of agents with different mobility preferences (such as different species or social classes). This way, each layer represents the mobility network of each type of agent, while each subpopulation is represented in each layer.

In an attempt to increase the realism of epidemiological models without compromising the possibility of a theoretical analysis, here we propose a mathematical framework in which the dynamical variables of each patch forming the metapopulation are treated independently. Our framework can accommodate any mobility multiplex network from real commuting datasets containing different types of individuals and is amenable to any particular distribution of the population across the patches, then generalizing previous findings on monolayer networks \cite{natphys, scarpino}. We will analyze the classical {\it Susceptible}-{\it Infected}-{\it Susceptible} (SIS) and the {\it Susceptible}-{\it Infected}-{\it Recovered} (SIR) models, achieving an excellent agreement with intensive Monte Carlo simulations. In addition, we derive an exact expression of the epidemic threshold and show its nontrivial dependence with the different mobility patterns represented in the multiplex. 

\section{Metapopulation model}

For the sake of clarity, we start considering a metapopulation framework consisting of one single type of agents. In this case we have a network composed of $N$ nodes (the patches) and a total population of $P$ agents. Importantly, each agent has associated one of the patches so that all the movements of agents associated to a node, say $i$, initiate from and return to it.  In its turn, a node $i$ of the network is the basement (or home) of a number $n_i$ of agents so that $P=\sum_{i=1}^{N} n_i$. For the sake of  generality, we consider that the network connecting the patches of the population is a weighted and directed graph, encoded in an adjacency matrix whose entries $W_{ij}$ account for the weight of the interaction from node $i$ to node $j$.

The dynamical model implemented in our metapopulation involves three different stages at each time step $t$: {\it movement}, {\it interaction} and {\it return} (MIR). First, each agent decides whether moving with probability $p$ or remaining in its associated home node $i$ with probability $(1-p)$. If the agent moves, it goes to any of the nodes connected to $i$ as dictated by the adjacency matrix ${\bf W}$. The probability that a patch $j$ is chosen, is proportional to the weight of the corresponding entry $W_{ij}$ of the adjacency matrix:
\begin{equation}
R_{ij}=\frac{W_{ij}}{\sum_{j=1}^{N}W_{ij}}\;.
\label{matrixR}
\end{equation}
Once all the agents have been placed in the nodes, the interaction stage takes place. Each agent updates its dynamical state according to the epidemic model at work (see below) by interacting with the agents that are placed in the same patch at time $t$.  Finally, agents come back to their corresponding residence node and another time step starts. These stages are depicted in Fig. \ref{fig:MIR} where, for clarity, we have considered that the states of agents are either {\it Healthy} or {\it Infectious} as in the SIS model.

This metapopulation model captures the commuting nature of most of human displacements within cities (at the level of neighborhoods) or countries (at the level of cities). Interestingly, let us remark that empirical data about real recurrent mobility patterns can be incorporated straightforward in the MIR model by considering the number of observed trips between two locations $W_{ij}$ in order to construct the transition rates matrix ${\bf R}$. This way, the model has as control parameters the displacement probability $p$ and those controlling the epidemic model under study.

\section{Population-based Markovian dynamics in Complex Networks}

In the following we will focus on the two most paradigmatic epidemic models, SIS and SIR. The reaction laws of these models are given by two parameters: {\it (i)} the probability $\lambda$ that a Susceptible (healthy) agent catches the diseases after the contact with a single infected individual and {\it (ii)} the probability $\mu$ that an infected overcomes the disease and turns to be susceptible again (SIS) or becomes immunized (SIR). These reactions can be expressed as: 
\begin{equation}
S+I\xrightarrow{\lambda}2I,\quad I\xrightarrow{\mu}S\;,
\label{eq:sis}
\end{equation}
for the SIS model, and:
\begin{equation}
S+I\xrightarrow{\lambda}2I,\quad I\xrightarrow{\mu}R\;,
\label{eq:sir}
\end{equation}
for the SIR.

As in any metapopulation model on large complex networks, we face the problem of computationally expensive simulations. A useful avenue to analyze these models, with the byproduct of obtaining analytical estimations for the impact of the epidemic, is to formulate coarse-grained models that reduce significantly the complexity of the problem. Typically, heterogeneous mean field (HMF) techniques have been applied in a number of works related to epidemic spreading in contact networks and metapopulations. As anticipated above, the main assumption of HMF is to correlate the relevant parameters of nodes and patches with their number of connections to other nodes, {\it i.e.} their degree. This way, two distant patches that are connected to the same number (but not the same set) of locations are considered to have the same static and dynamical properties such as, for instance, the number of habitants and the fraction of infected agents. This assumption, although being strong, has been shown to be valid for small epidemic sizes, thus allowing quite good predictions of epidemic thresholds.

Here we formulate the mathematical equations of the MIR model by  following a similar avenue as in \cite{EPL,Guerra,PRE} for contact networks, thus generalizing the Markovian approach to complex metapopulations. This way, we will consider both static and dynamical variables of each individual patch as independent, allowing us to compare directly with the findings of Monte Carlo simulations at the microscopic level and, more importantly, to derive theoretical results for any kind of particular mobility networks.
 
\subsection{SIS model}

For the SIS model, we have a set of $N$ variables $\rho_i(t)$ denoting the fraction of infected agents associated to patch $i$ at time $t$. It is important to stress that, according to the MIR model, an agent whose associated patch is $i$ can be in other node $j$ at time $t$. The time evolution of $\rho_i(t)$ can be written as: 
\begin{equation}
\rho_i(t+1)=(1-\mu)\rho_i(t)+(1-\rho_i(t))\Pi_{i}(t)\;,
\label{sis}
\end{equation}
where the first term denotes the fraction of infected agents associated to $i$ that do not recover at time $t+1$. The second term instead accounts for the fraction of healthy agents associated to $i$ that pass to infected at time $t+1$. In this second term, $\Pi_i(t)$ is the probability that a healthy agent associated to node $i$ becomes infected at time $t$. This probability reads:
\begin{equation}
\Pi_i(t)=(1-p)P_i(t)+p\sum_{j=1}^{N}R_{ij}P_j(t)\;,
\label{Eq:probSIS}
\end{equation}
where the first term denotes the probability that a susceptible agent associated to patch $i$ becomes infected when remaining at its home node $i$ and the second one accounts for the probability that this agent catches the disease when moving to any neighbor of $i$. 

Finally, the probability $P_i(t)$ in Eq. (\ref{Eq:probSIS}) denotes the probability that a healthy agent in (but not necessarily associated to) node $i$ at time $t$ becomes infected after the contact with any of the infected agents present inside $i$ at the same time. Then, probability $P_i(t)$ reads:
\begin{equation}
P_i(t)=1-\prod_{j=1}^{N}(1-\lambda \rho_{j}(t))^{n_{j\rightarrow i}}\;
\label{probinf}
\end{equation}
where:
\begin{equation}
n_{j\rightarrow i}=\delta_{ij}(1-p)n_i+pR_{ji}n_j\;,
\label{effpops}
\end{equation}
being $\delta_{ij}=1$ when $i=j$ and $\delta_{ij}=0$ otherwise. 

\begin{figure}[t!]
\begin{center}
\includegraphics[width=0.60\columnwidth,angle=-90]{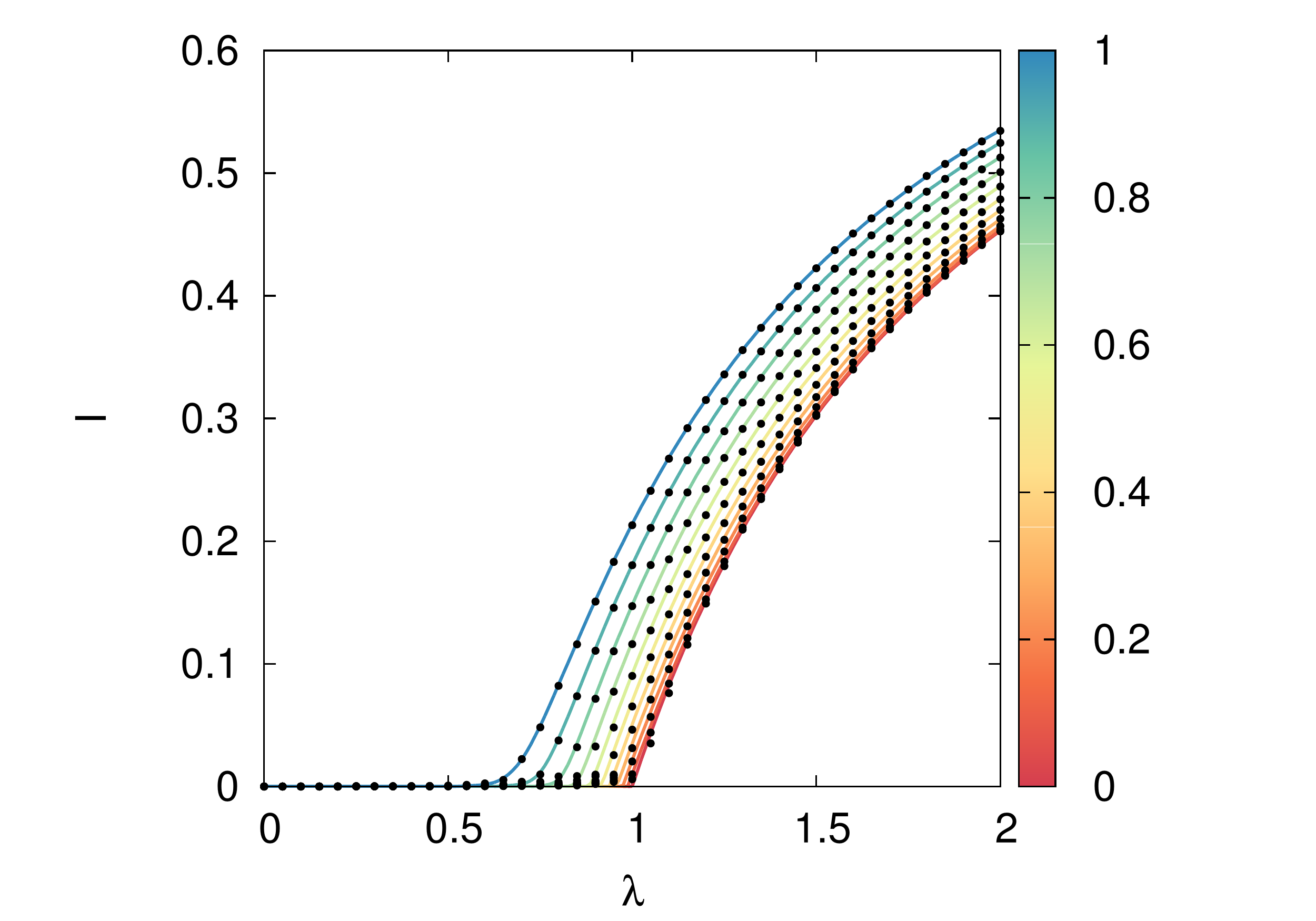}
\includegraphics[width=0.60\columnwidth,angle=-90]{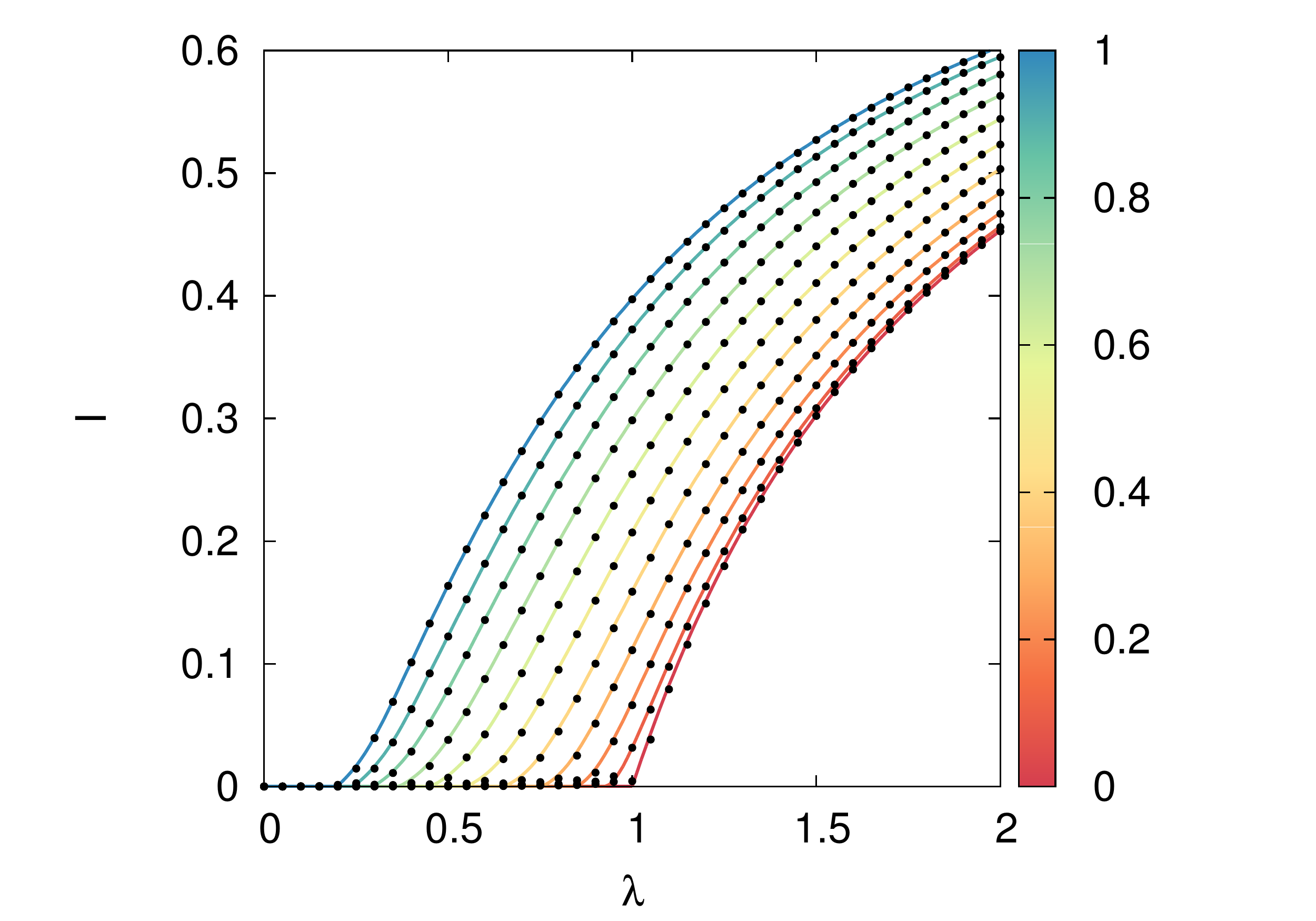}
\caption{\label{figSIS} (color online) $I(\lambda)$ for the SIS dynamics in ER (top)  and SF (bottom) networks of $10^3$ nodes and $\langle k\rangle=5.5$ and $\langle k\rangle=7.3$ respectively. The population of each node is $5\cdot 10^3$ individuals.The solid curves indicate the solution obtained by solving the Markovian evolution equations (the color of each curve indicates the value of $p$ as shown in the color bars),  whereas the points correspond to the results obtained by using MC simulations ($20$ realizations for each value of $\lambda$). Note that the value of $\lambda$ has been re-scaled by the critical value at $p=0$, {\it i.e.}, that of a well-mixed population of $n=5000$ individuals: $\lambda_c(p=0)=\mu/n=4\cdot 10^{-4}$. The recovery rate is $\mu=0.2$. 
}
\end{center}
\end{figure}

The expressions in Eqs. (\ref{sis})-({\ref{effpops}) compose the closed set of equations covering the evolution of an SIS disease spreading in the MIR metapopulation model with parameters $p$, $\mu$ and $\lambda$. In addition, matrix ${\bf R}$ is given by the topology of the mobility network, that can be constructed from the observed flows between the patches, and the set of node populations, $\{n_i\}$, can be also set according to the local census of the population under study.

\subsection{SIR dynamics}
  
The formulation of the Markovian equations for a metapopulation under a SIR spreading dynamics demands to add another set of $N$ variables: $\{r_{i}(t)\}$ ($i=1,...,N$), {\it i.e.}, the fraction of recovered agents associated to patch $i$. Thus, the set of $N$ equations (\ref{sis}) for the SIS model is now substituted by the following set of $2\cdot N$ equations:
\begin{eqnarray}
\rho_i(t+1)&=&(1-\mu)\rho_i(t)+(1-\rho_i(t)-r_i(t))\Pi_{i}(t)\;,
\label{SIR1}
\\
r_i(t+1)&=&r_i(t)+\mu\rho_i(t)\;,
\label{SIR2}
\end{eqnarray}
On the other hand, since the infection processes within each of the patches in the SIR model follow identical rules as those of the SIS one, the expression in Eq. (\ref{SIR1}) for the probability that a healthy agent associated to node $i$ becomes infected at time $t$, $\Pi_{i}(t)$, has the same form as in the SIS case. This way, the SIR metapopulation dynamics is fully described by Eqs. (\ref{SIR1}) and (\ref{SIR2}) with the addition of Eqs. (\ref{Eq:probSIS})-(\ref{effpops}).

\begin{figure}[t!]
\begin{center}
\includegraphics[width=0.60\columnwidth,angle=-90]{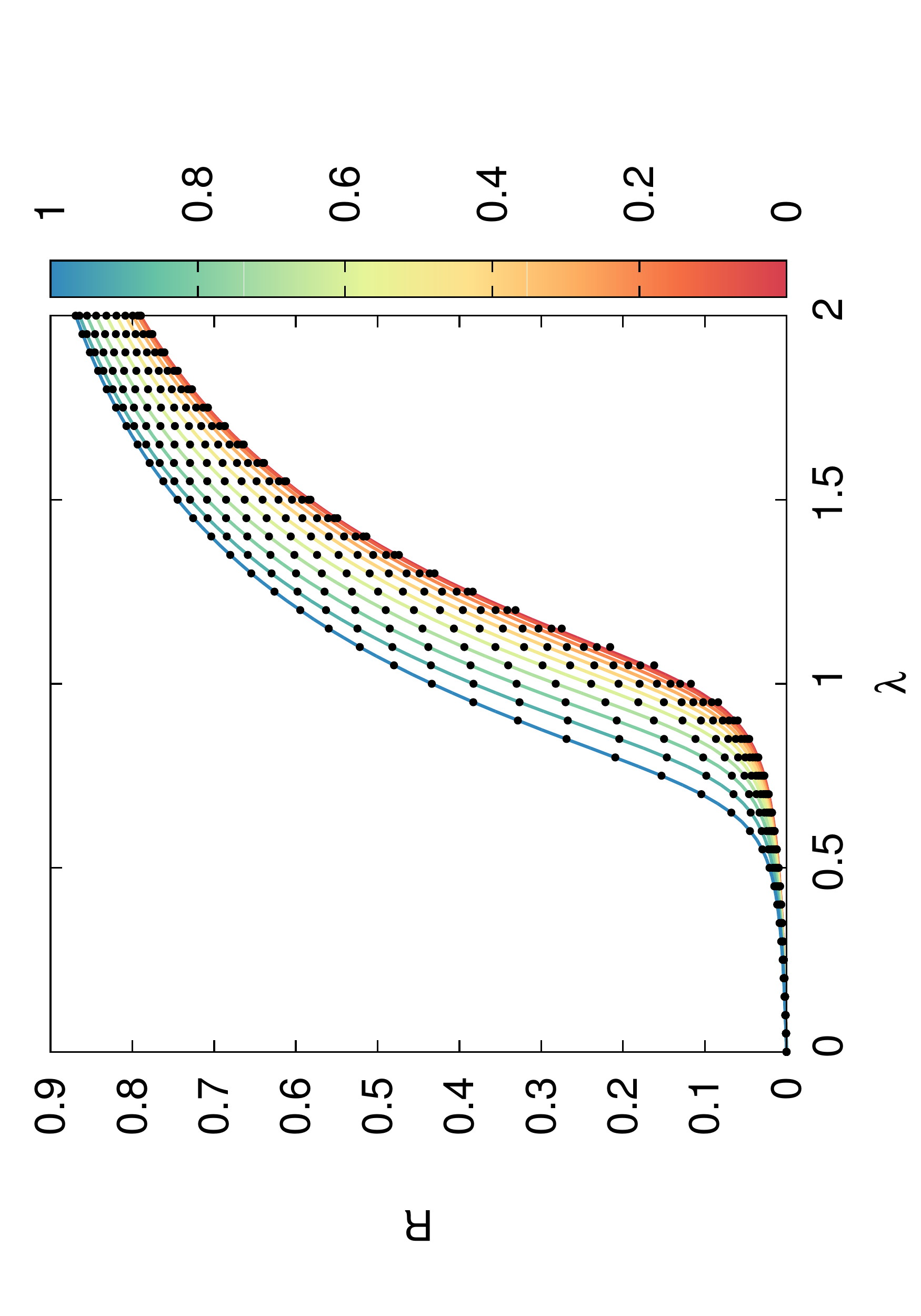}
\includegraphics[width=0.60\columnwidth,angle=-90]{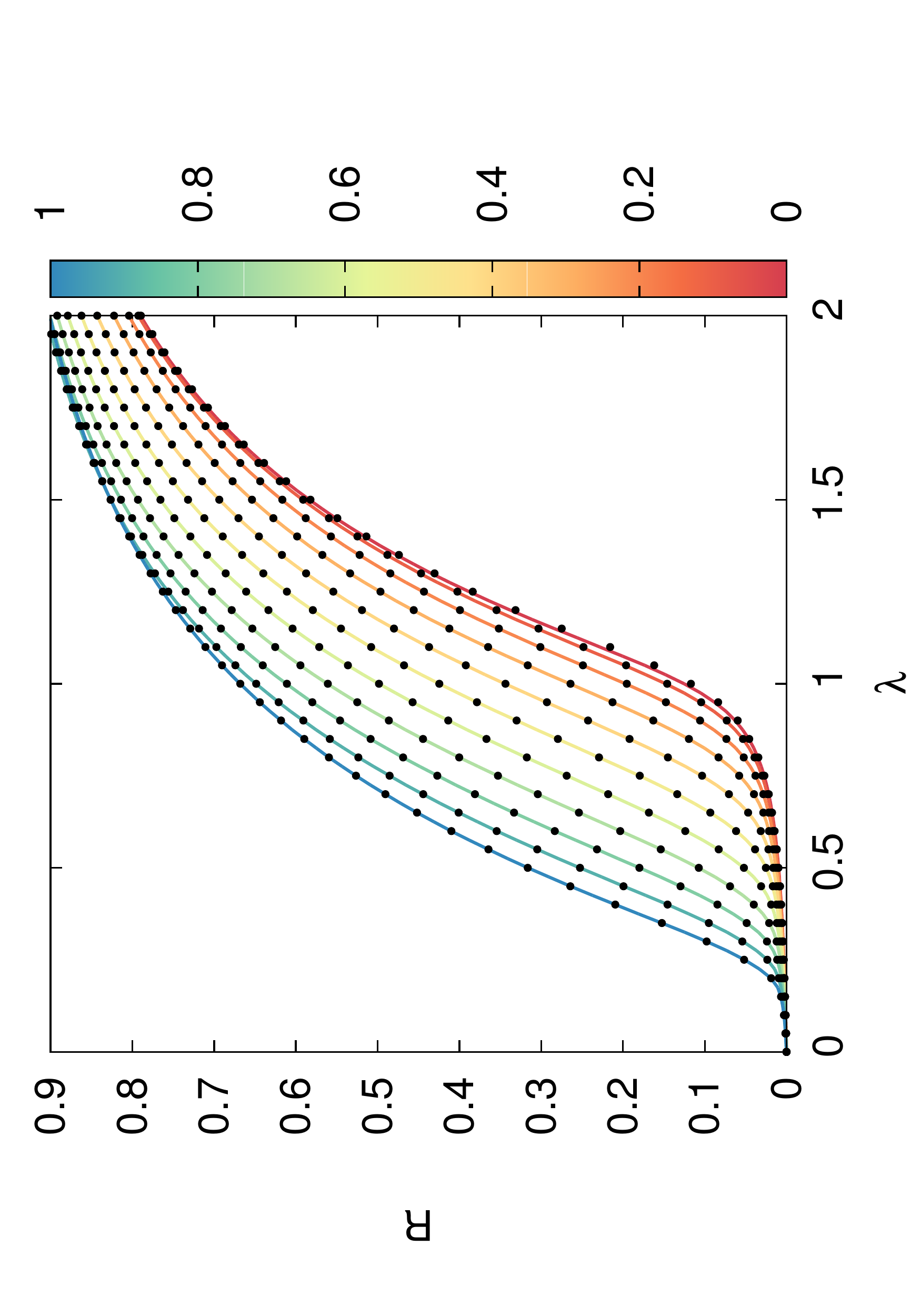}
\caption{\label{figSIR} $R(\lambda)$ for the SIR dynamics in ER (top)  and SF (bottom) networks of $10^3$ nodes and $\langle k\rangle=5.5$ and $\langle k\rangle=7.3$ respectively. The population of each node is 5000 individuals.The solid curves indicate the solution obtained by solving the Markovian evolution equations (the color of each curve indicates the value of $p$ as shown in the color bars),  whereas the points correspond to the results obtained by using MC simulations ($10^2$ realizations for each value of $\lambda$). Note that the value of $\lambda$ has been re-scaled by the critical value at $p=0$, {\it i.e.}, that of a well-mixed population of $n=5\cdot 10^3$ individuals: $\lambda_c(p=0)=\mu/n=4\cdot 10^{-4}$. The recovery rate is $\mu=0.2$.
}
\end{center}
\end{figure}

\subsection{Validation of the Markovian equations}

To check the accuracy of the Markovian equations we have considered Erd\"os-R\'enyi (ER) and Scale-free (SF) synthetic networks having the same number of nodes $N=10^3$ and average connectivities $\langle k\rangle=5.5$ and  $\langle k\rangle=7.3$ respectively. The nodes of these networks are homogeneously populated, $n_i=5\cdot 10^3$ $\forall i$, so that the total population of our systems is $P=5\cdot 10^6$. The weights $W_{ij}$ between the nodes of the graphs are randomly assigned following a homogeneous distribution within the range $W_{ij}\in [1,50]$. Once all the weights are set, we construct the transition matrix ${\bf R}$ [see Eq. (\ref{matrixR})] for each graph.

Monte Carlo simulations start by infecting a small fraction of agents in each of the nodes. In particular, we infect each agent with probability $10^{-3}$ so that, on average, there is $1$ infected agent per node at time $t=0$.  This initial configuration corresponds to set as initial conditions of the Markovian equations $\rho_i(0)=10^{-3}$ $\forall i$ (and $r_i(0)=0$ $\forall i$ in the SIR case). For Monte Carlo simulations, due to the stochastic nature of the initial configuration and the disease models, we have averaged the results over $10^2$ realizations for each combination of the parameters ($p$, $\lambda$ and $\mu$) considered.

First we analyze the SIS model. In Fig. \ref{figSIS} (top and bottom panels correspond to ER and SF networks respectively) we plot the number of infected agents in the steady state, $I$, as a function of the infection probability, $\lambda$, for different movement probabilities $p$. The points denote the results of Monte Carlo simulations for each value of $\lambda$ and $p$ while solid curves correspond to the solution of the Markovian equations. The agreement between simulations and the equations is almost exact, capturing with high accuracy the macroscopic state of the metapopulation both in the disease-free and epidemic regimes. For the SIR model in Fig. \ref{figSIR}  (top and bottom panels correspond to ER and SF networks respectively)  we plot the number of recovered agents, $R$, as function of $\lambda$ for the same set of movement probabilities $p$ as for the SIS model. Again, we observe the exact agreement between Monte Carlo simulations and the solution of Markovian equations both before and after the epidemic threshold.

\begin{figure}[t!]
\begin{center}
\includegraphics[width=0.95\columnwidth,angle=-0]{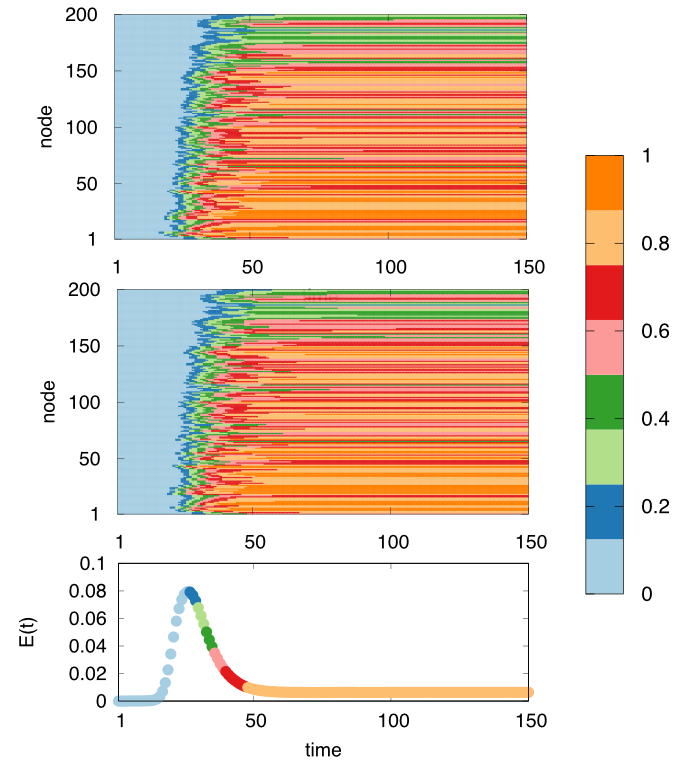}
\caption{Microscopic evaluation of the Markovian equation. The top panel shows the time evolution, according to the agent based simulation, of the fraction of recovered agents of each node, $r_i(t)$ in an SF network of $N=200$ when a initial infectious seed is placed at a single node. The population of the system is $P=7\cdot 10^5$ agents while the size of a node is proportional to its out strength (see text). The weights of the links are randomly assigned following an homogeneous distribution in the range $W_{ij}\in[1,50]$. The panel in the middle of the figure shows the same scenario but solved using the Markovian equations. Finally, to evaluate the accuracy of the Markovian equations we plot in the bottom panel the time evolution of the error [see Eq. (\ref{error})] with respect to the Monte Carlo simulation. The parameters used are: $p=0.1$, $\lambda=2\lambda_c(p=0)$ and $\mu=0.2$.
\label{figEvol}
}
\end{center}
\end{figure}

The high accuracy of the solution of Markovian equations shown in Figs. \ref{figSIS} and \ref{figSIR} allows us to overcome the computational costs associated to large scale Monte Carlo simulations. However, it is clear that both $I$ (SIS) and $R$ (SIR) are macroscopic indicators of the outreach of the disease in the whole population. The examples shown above assume that the populations across nodes are homogeneously distributed. However, in real metapopulations, such as cities, each patch contains a different number of agents. These demographic heterogeneity may lead to interesting effects, such as the increase of the epidemic threshold with the increase of mobility \cite{natphys,scarpino}. Here, due to the homogenous distribution of agent across patches, mobility always leads to a decrease of the epidemic onset (as shown in Figs. \ref{figSIS} and \ref{figSIR}).

To check further the accuracy of the Markovian equations we now increase the level of resolution and monitor the spatio-temporal evolution of the infections in the population. To this aim we now fix $\lambda$, $\mu$ and $p$, and set a small infected fraction of agents placed at the same node. Then we run an SIR epidemics and follow how this infectious seed spreads across the patches of the metapopulation by monitoring $r_i(t)$. 

The results of the agent based simulation are shown in the top panel in Fig. \ref{figEvol} where we have used an SF network of $N=200$ nodes and a total population of $P=7\cdot 10^5$ agents where each patch contains a number of agents proportional to its out-strength ($s_{i}^{out}=\sum_{j=1}^{N}W_{ij}$). We have ranked the nodes (from $1$ to $200$) according to the order in which infections occur, so that node $1$ is the one in which the initial infectious seed is placed. As shown, after a transient time of roughly $50$ time steps the system reaches the frozen state and the density of recovered agents at each patch remains stationary. Interestingly, the final pattern clearly shows that the local significance of the disease is not homogeneous. 

The solution of the Markovian evolution equations (second panel of Fig. \ref{figEvol}) fairly reproduces the spatio-temporal pattern  from Monte Carlo simulations. To quantify the agreement, we have computed the error in quantifying the number of recovered and infected agents by the Markovian equations per time step as:
\begin{equation}
E(t)=\frac{1}{N}\sum_{i=1}^{N}\left|\left(r_i(t)+\rho_i(t)\right)-\frac{\left(R_i(t)+I_i(t)\right)}{n_i}\right|\;,\\
\label{error}
\end{equation}
where $R_i(t)$ and $I_i(t)$ are, respectively, the number of recovered and infected agents associated to node $i$ at time $t$ observed in Monte Carlo simulation while $r_i(t)$ and $\rho_i(t)$ are, respectively, the fractions of recovered and infected individuals as obtained when solving the Markovian equations. The bottom panel in Fig. \ref{figEvol} shows that the Error reaches its maximum value ($8\%$) just after the avalanche of contagions across the metapopulation starts. After this peak, $E(t)$ reaches a stationary value around $1\%$ pointing out the high accuracy in reproducing the stationary pattern shown above. 

\section{Multiplex Metapopulations}

We address now the case of metapopulations in which different types of agents interplay. In particular we will focus on systems in which agents displaying $L$ types of mobility patterns coexist within each patch. This way, the population of a patch $i$ is the sum of the number of agents of each type $n_i=\sum_{\alpha=1}^{L} n_{i}^{\alpha}$ and the probability that an agent of patch $i$ and type $\alpha$ visits another patch $j$ is now written as the generalization of Eq. (\ref{matrixR}):
\begin{equation}
R_{ij}^{\alpha}=\frac{W_{ij}^{\alpha}}{\sum_{j=1}^{N}W_{ij}^{\alpha}}\;,
\end{equation}
where $W_{ij}^{\alpha}$ is associated to the number of observed trips of agents of type $\alpha$ in patch $i$ to patch $j$.

To analyze this situation, it is natural to make use of a multiplex formulation \cite{kivela2014,boccaletti2014,dedomenico2013,battiston,radicchi} of the metapopulation, as it is illustrated in Fig. \ref{fig:multiplex}. In our case, the number of layers of the multiplex is equal to the number of types of agents ($L$) and the architecture of each layer is described by a different matrix ${\bf R}^{\alpha}$. Each patch of the system is represented as one node in each network layer and the corresponding $L$ nodes are virtually connected (dotted lines) as they mix their agents when the contagion processes take place.

\begin{figure}[b!]
\begin{center}
\includegraphics[width=0.92\columnwidth,angle=-0]{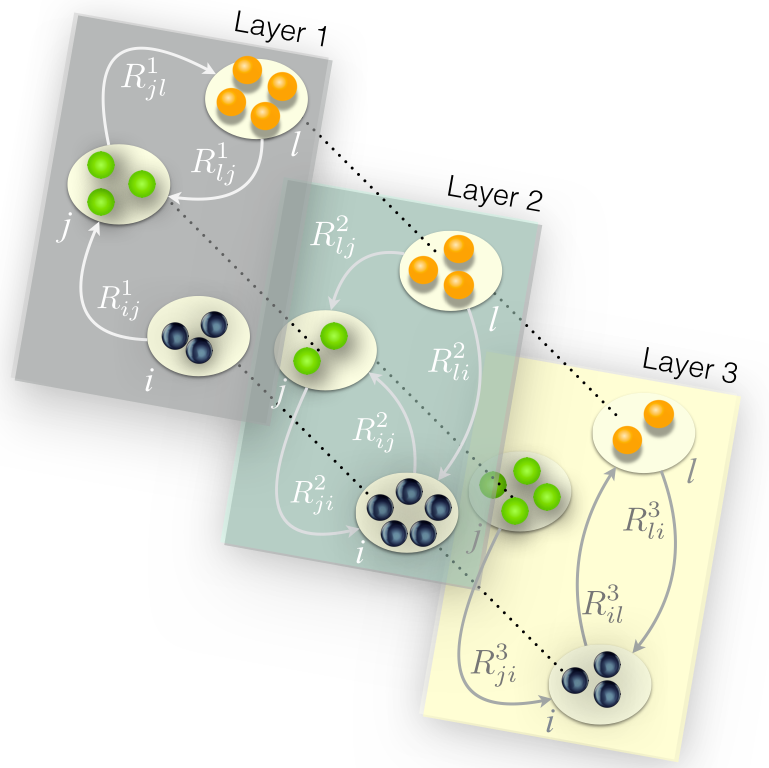}
\caption{\label{fig:multiplex} Schematic representation of a metapopulation multiplex composed of $L=3$ layers. Each of the $N=3$ patches (nodes) is represented in each of the layers. The layers highlight that individuals of type $\alpha$ associated to patch $i$ move to another patch $j$ with probability $R_{ij}^{\alpha}$ which, in general, is different from the rate of transitions of agents of type $\beta\neq\alpha$ associated to the same node. This way each layer $\alpha$ presents a topology captured by a different matrix ${\bf R^{\alpha}}$. 
}\end{center}
\end{figure}

The number of Markovian equations of the multiplex are now multiplied by $L$ with respect to the networked metapopulation. In particular for the SIS (and SIR) model, the variables are $\rho_{i}^{\alpha}(t)$ (and $r_i^{\alpha}(t)$), which denote the fraction of infected (and recovered) individuals of layer $\alpha=1,...,L$ associated to node $i$. In this case, SIR equations become:
\begin{widetext}
\begin{eqnarray}
\rho^{\alpha}_i(t+1)&=&\rho^{\alpha}_i(t)(1-\mu)+
\left(1-\rho^{\alpha}_i(t)-r^{\alpha}_i(t)\right)\left((1-p)P^{\alpha}_i(t) + p\sum_{j=1}^{N}R^{\alpha}_{ij} P^{\alpha}_{j}(t)\right)\;, 
\label{eqrho}\\
r^{\alpha}_i(t+1)&=&r^{\alpha}_i(t)+(1-\mu)\rho^{\alpha}_i(t)\;,
\end{eqnarray}
\end{widetext}
while for the SIS model we only have Eq. (\ref{eqrho}) with $r_{i}^{\alpha}(t)=0$. The term $P^{\alpha}_i(t)$, which denotes the probability that an agent of type $\alpha$ placed in patch $i$ at time $t$ becomes infected, reads: 
\begin{equation}
P^{\alpha}_{i}(t)=1-\prod_{\beta=1}^{L}\prod_{j=1}^{N}\left(1-\lambda^{\beta\alpha}\rho^{\beta}_j(t)\right)^{n^{\beta}_{j\rightarrow i}(t)}
\label{multiplexP}
\end{equation}
where $\lambda^{\beta\alpha}$ is the probability that an diseased agent of type $\beta$ infects a healthy agent of type $\alpha$. In addition the number of agents of type $\alpha$ associated to patch $j$ that travel to a different patch $i$ is given by:
\begin{equation}
n_{j\rightarrow i}^{\alpha}=(1-p)\delta_{ij}n_{i}^{\alpha}+pR_{ji}^{\alpha}n_{j}^{\alpha}\;.
\label{travels}
\end{equation}
The set of Eqs. (\ref{eqrho})-(\ref{travels}) conforms the Markovian model of  the multiplex metapopulation. For the sake of simplicity, we will now restrict to the case $\lambda^{\alpha\beta}=\lambda$ $\forall \alpha ,\beta$, so that the infection probability between healthy and infected agents does not depend on their types.

\begin{figure*}[t!]
\begin{center}
\includegraphics[width=0.47\columnwidth,angle=-90]{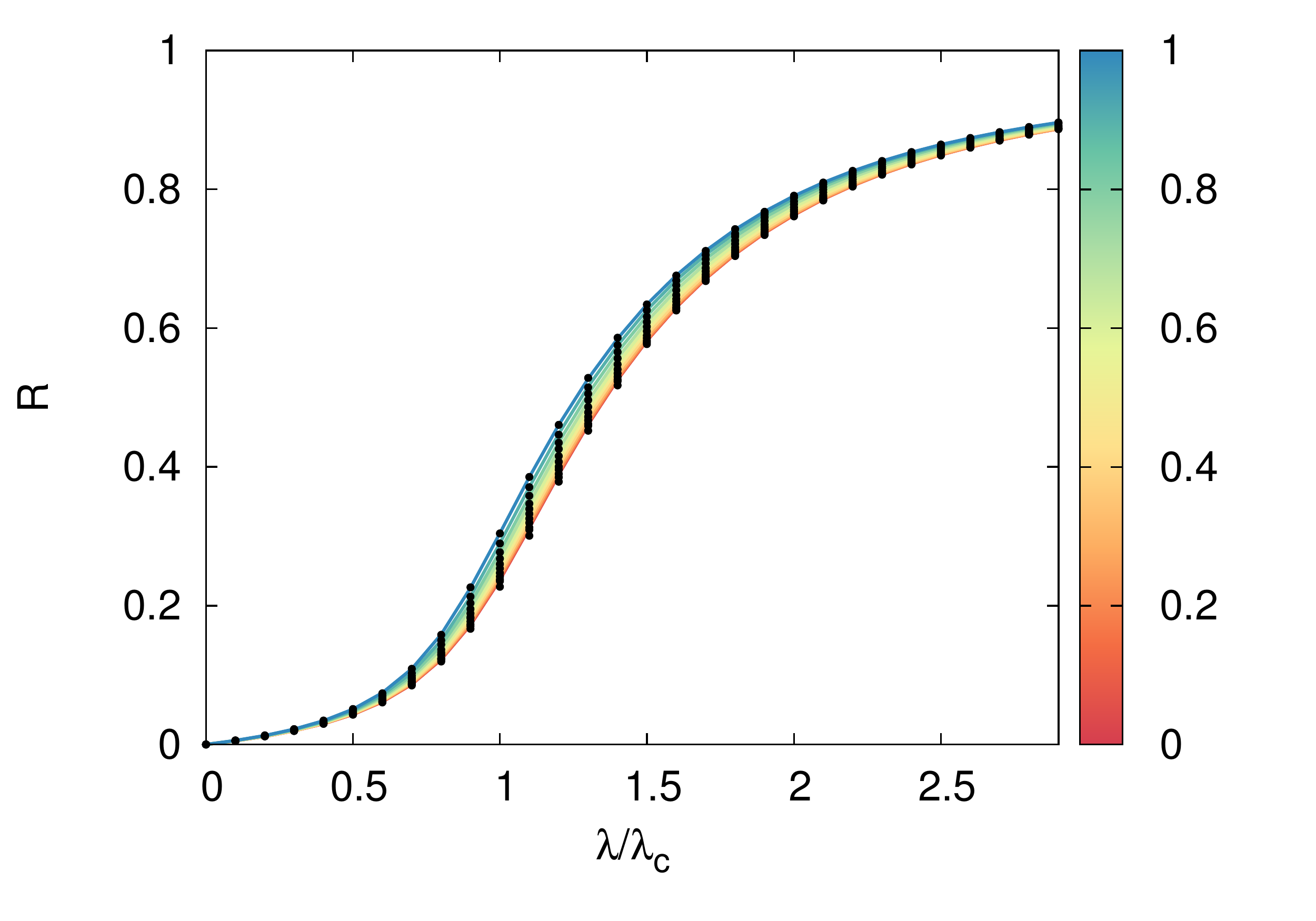}
\includegraphics[width=0.47\columnwidth,angle=-90]{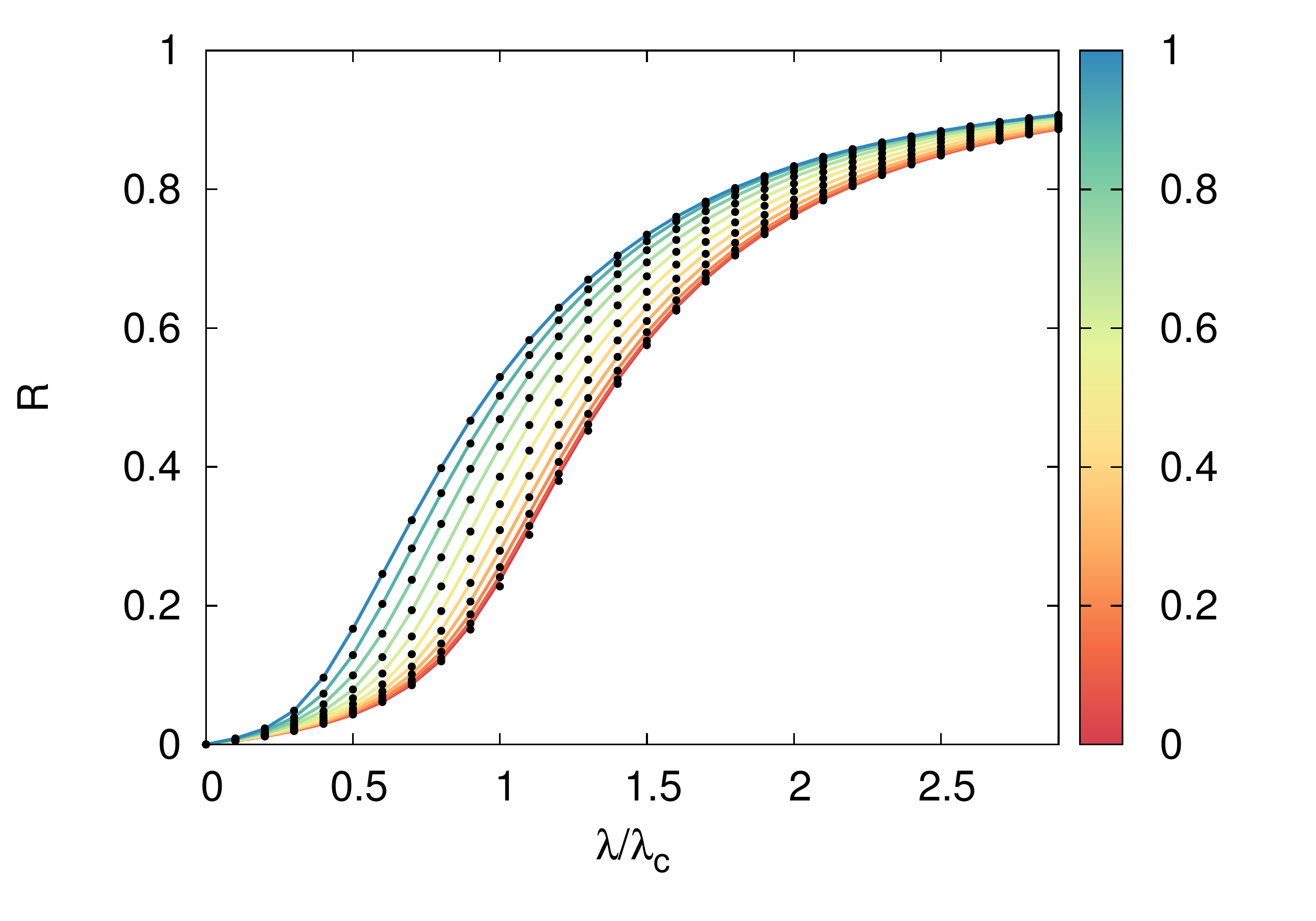}
\includegraphics[width=0.47\columnwidth,angle=-90]{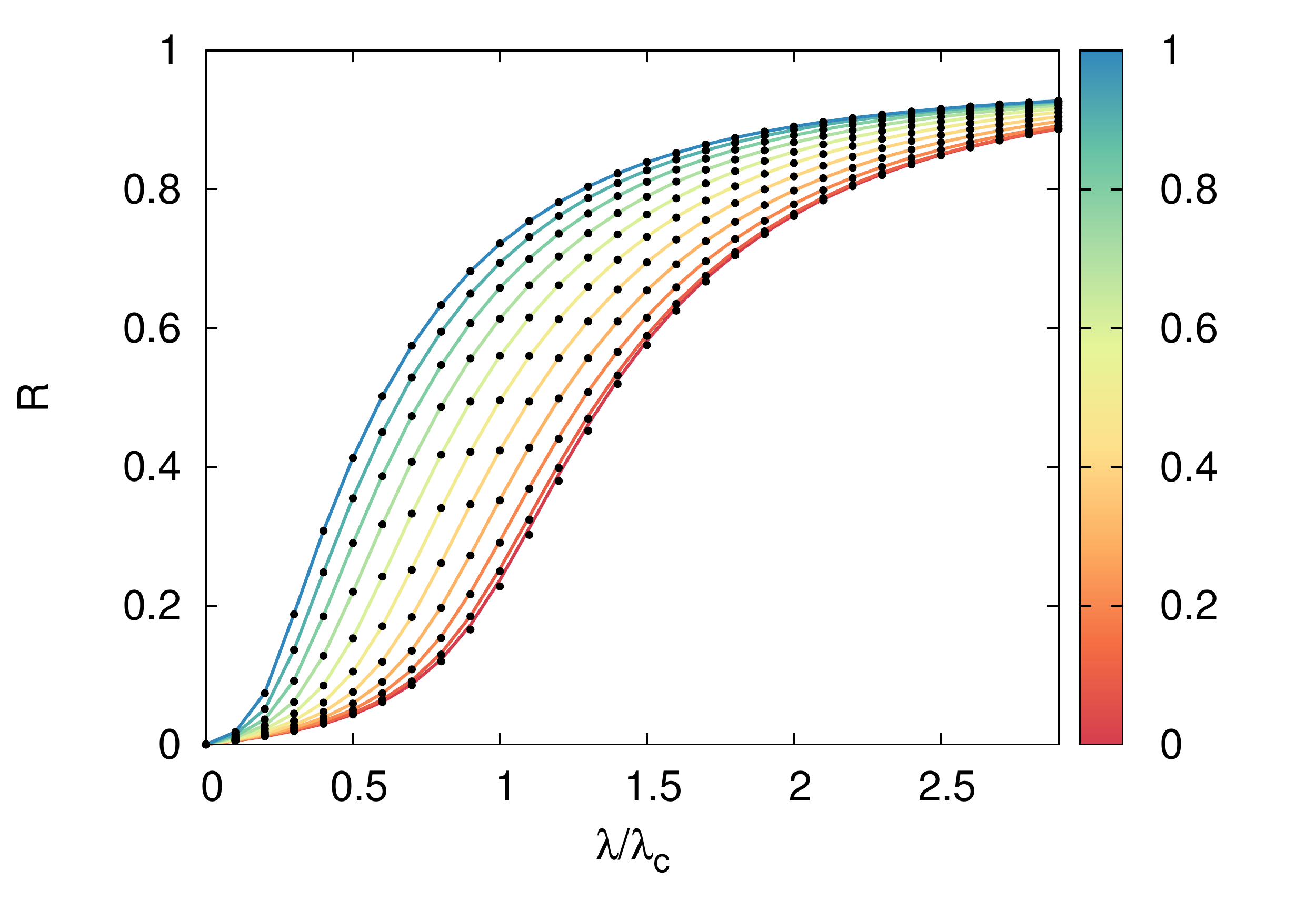}
\includegraphics[width=0.47\columnwidth,angle=-90]{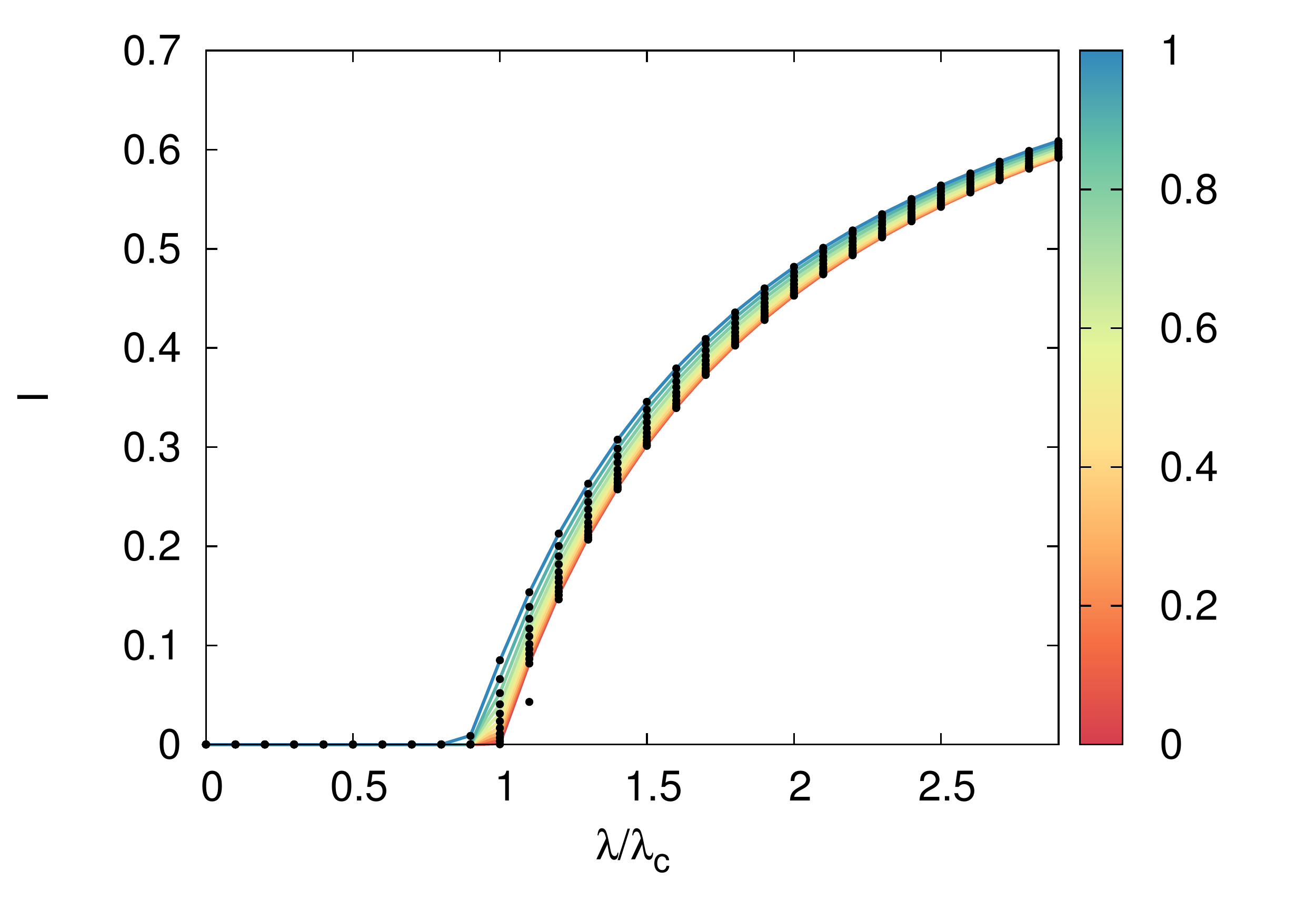}
\includegraphics[width=0.47\columnwidth,angle=-90]{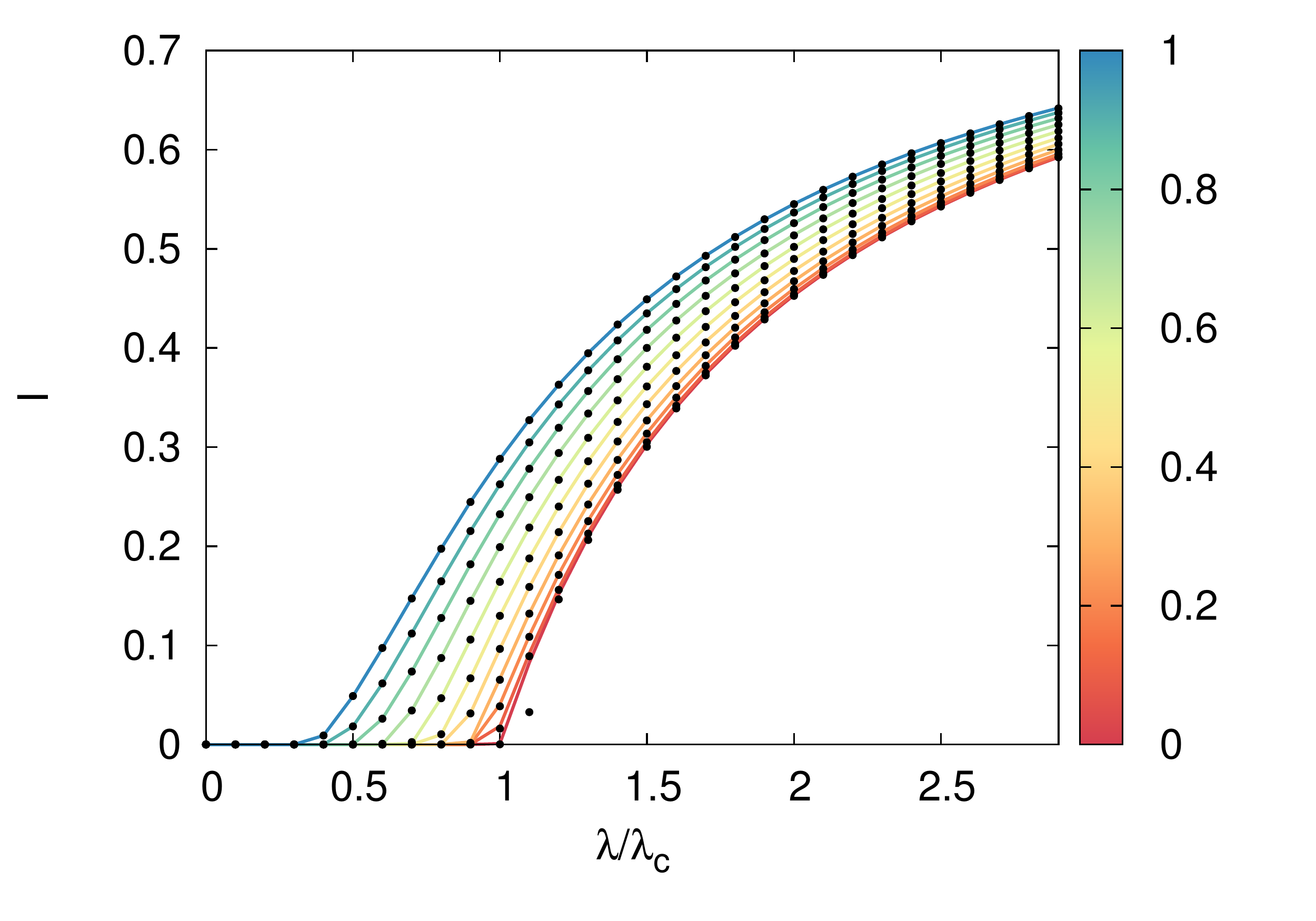}
\includegraphics[width=0.47\columnwidth,angle=-90]{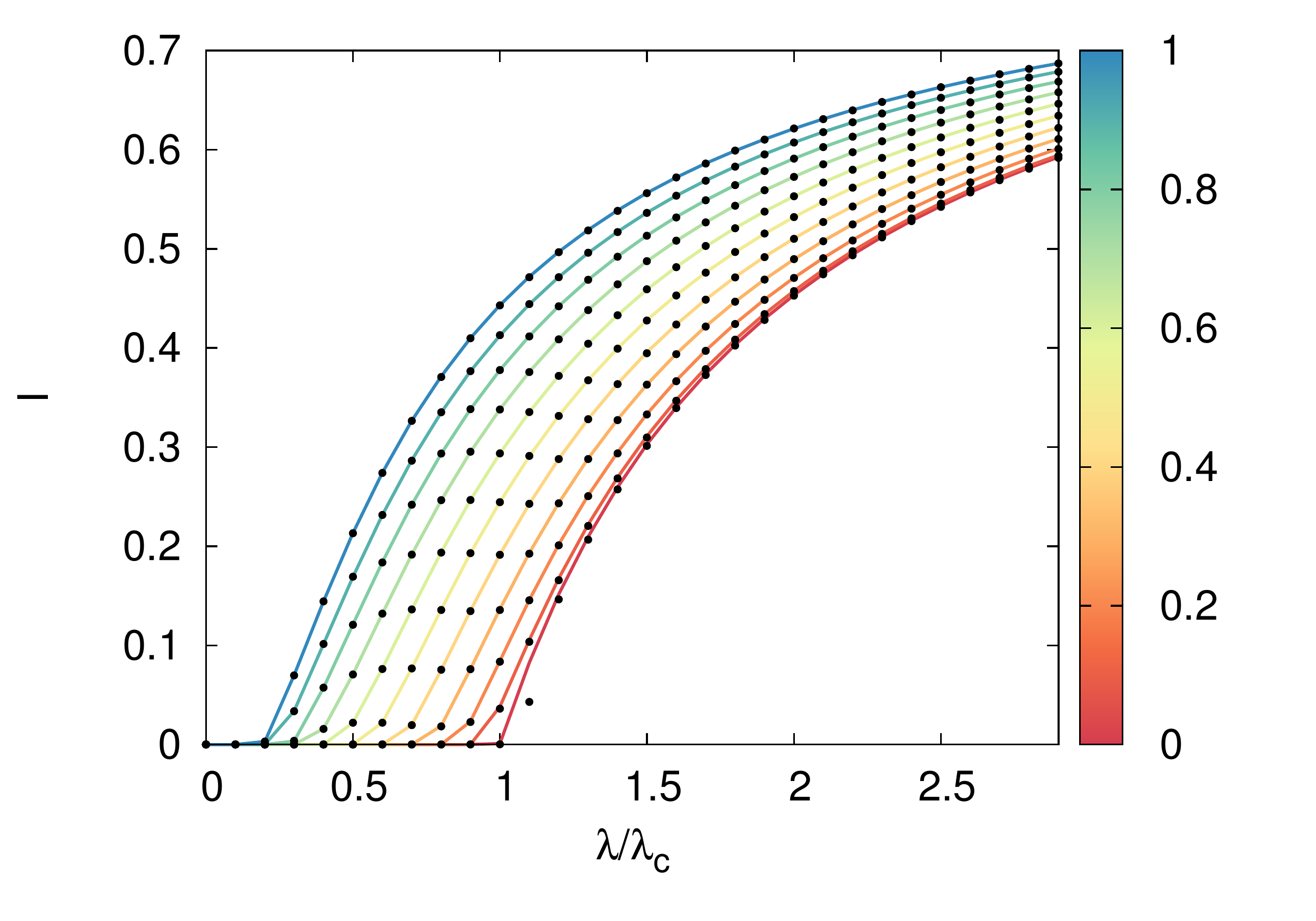}
\caption{\label{fig:markovmultiplex} Epidemic diagrams for the SIR (top), $R(\lambda)$ and SIS (bottom), $I(\lambda)$ dynamics of three different multiplexes with $L=2$ layers. From left to right we have ER-ER, ER-SF, and SF-SF. In all the cases each network layer has $N=10^3$ nodes and each node contains $500$ individuals per layer. The solid curves indicate the solution obtained by solving the Markovian evolution equations (the color of each curve indicates the value of $p$ as shown in the color bars),  whereas the points correspond to the results obtained by using agent based simulations ($50$ realizations for each value of $\lambda$ and $p$). Note that the value of $\lambda$ has been re-scaled by the critical value $\lambda_c$ at $p=0$, {\it i.e.}, that of a well-mixed population of $n= 10^3$ individuals: $\lambda_c =\mu/10^{3}$ at $p=0$. The recovery rate is $\mu=0.2$
}
\end{center}
\end{figure*}

\subsection{Validation or the Markovian equations}

To validate the Markovian equations for the multiplex metapopulation we proceed in the same fashion as we did for networked ones. First, we compute the impact that SIR and SIS diseases have as a function of the infectivity of the disease, $\lambda$, and  the degree of mobility, $p$. We have studied three types of multiplex of $L=2$ layers, namely ER-ER, SF-SF, and ER-SF, of $N=10^3$ nodes and each node has an identical population of $500$ agents. The weights of each link $W_{ij}^{\alpha}$ is randomly assigned following an homogeneous distribution in the range $[1,50]$.

In Fig. \ref{fig:markovmultiplex} we show the diagrams for the SIR (top) and the SIS (bottom) where dots represent the results obtained for Monte Carlo simulations of the epidemic processes and the solid lines are for the solution of the Markovian equations. As in the case of networked metapopulations, we observe a perfect agreement between simulations and the numerical solution of Eqs. (\ref{eqrho})-(\ref{travels}). From the physical point of view we observe that, while for all the cases mobility enhances the anticipation of the epidemic onset, the multiplex composed of an ER and a SF topologies yields an intermediate anticipation effect compared to those observed for ER-ER and SF-SF. This is an interesting result that differentiates what has been recently observed in epidemic processes in multiplex contact networks \cite{cozzo,arruda}, where coupling $L$ layers yields an overall epidemic threshold that is equal to the smallest threshold of the isolated layers or, in other words, the epidemic onset is driven by the largest of the maximum eigenvalues of the set of adjacency matrices that define the layers. It is clear that the case of metapopulations the situation is more complicated as we show in the following section.

\begin{figure*}[t!]
\begin{center}
\includegraphics[width=1.95\columnwidth,angle=-0]{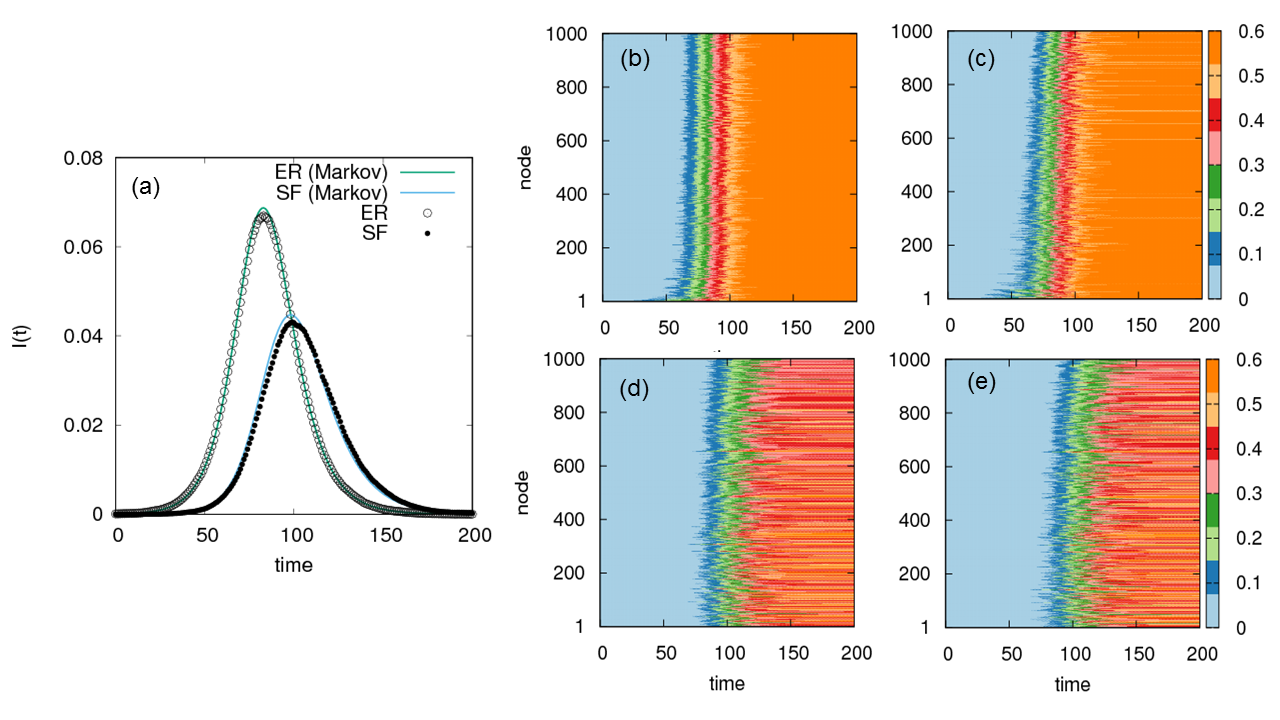}
\caption{Spatio-temporal patterns of the SIR dynamics in a metapopulation multiplex composed of an ER and a SF layer. Each layer has $10^3$ patches and $500$ individuals are associated to each patch. Theoretical prediction are shown by lines whereas dots represent Monte Carlo results. The initial infected agents are placed in a single patch of the ER layer. This, together with the small contagion probability between agents of different layers (see the text for details), causes the time difference between the epidemic onsets in each layer as observed from panel (a). Panels (b)-(d) show the time evolution of the fraction of recovered agents for each patch. The top panels show this evolution in the ER layer obtained for Monte Carlo simulations [panel (b)] and the solution of the Markovian model [panel (c)]. On the other hand, bottom panels show the same evolution in the SF layer as obtained again from simulations [panel (d)] and by solving the Markovian equations (\ref{eqrho})-(\ref{travels}) [panel (e)].
}\label{evolcapas}
\end{center}
\end{figure*}

We now focus on the general scenario in which $\lambda^{\alpha\beta}\neq \lambda$, {\it i.e.}, the contagion probability between two agents depends on their corresponding types. To this aim, we consider one population of agents whose movements are described by an ER mobility network and another population whose movements occur according to an SF graph. The number of patches is $N=10^3$ and inside each patch there are $500$ agents of each type (ER and SF). We consider the situation in which $\lambda^{\alpha\beta}\ll \lambda^{\alpha\alpha}$ ($\alpha\neq \beta)$. In particular contagion between agents moving in the ER layer occur with probability $\lambda^{ER}=1.5\mu/500$ and that for the agents moving in the SF layer is set to $\lambda^{SF}=1.1 \mu/500$ (recall that $\mu/500$ is the epidemic threshold for a well mixed population of $500$ agents). In its turn, we have set the infection probability between agents of different type to $\lambda^{ER-SF}=\lambda^{SF-ER}=0.025\mu/500$. Finally, to work with a more heterogeneous setup, we study the case of an SIR dynamics in which a small seed of initial infected agents is set in a single patch and affect only agents of one type (here those moving across the ER layer).

To analyze the accuracy of Eqs. (\ref{eqrho})-(\ref{travels}) in capturing the spatio-temporal evolution of epidemics, we first consider the temporal evolution of the fraction of infected individuals of each type (layer). In panel (a) of Fig.~\ref{evolcapas}, we show this evolution comparing the solution of the Markovian equations (solid lines) with the result obtained from Monte Carlo simulations (points). It is clear that the Markovian equations (\ref{eqrho})-(\ref{travels}) fairly reproduce the output of the numerical simulation, capturing  the delay of the onset of the epidemics in the SF layer with respect to that in the population moving across the ER. This delay is a clear consequence of {\it (i)} the localization of the initial infected individuals in the ER layer and {\it (ii)} the small contagion probability between agents of different type (layer). Interestingly, the fact that  $\lambda^{ER-SF}$ is far less than the threshold ($\mu/500$) in a closed population of $500$ agents does not prevent the disease from invading the  SF layer. 
 
Finally, in Fig.~\ref{evolcapas}.(b)-(e) we show the temporal evolution of the fraction of recovered individuals for each patch in each of the layers (ER top and SF bottom) obtained from numerical simulations (left panels) and solving Eqs. (\ref{eqrho})-(\ref{travels}) (right panels). The fair agreement between left and right panels indicates the great spatio-temporal accuracy of the Markovian model. Here, in addition to the delay in the onset of the epidemics in the SF population already observed in (a), it is remarkable that two different stationary regimes are obtained in each layer. Namely, the fraction of recovered individuals in the ER layer is nearly identical for all the patches. However, in the SF population the stationary pattern  points out a far more heterogeneous distribution of recovered individuals across the different patches. 

\subsection{Real Multiplex Metapopulations}
\begin{figure}[b!]
\centering
\includegraphics[width=1.05\columnwidth]{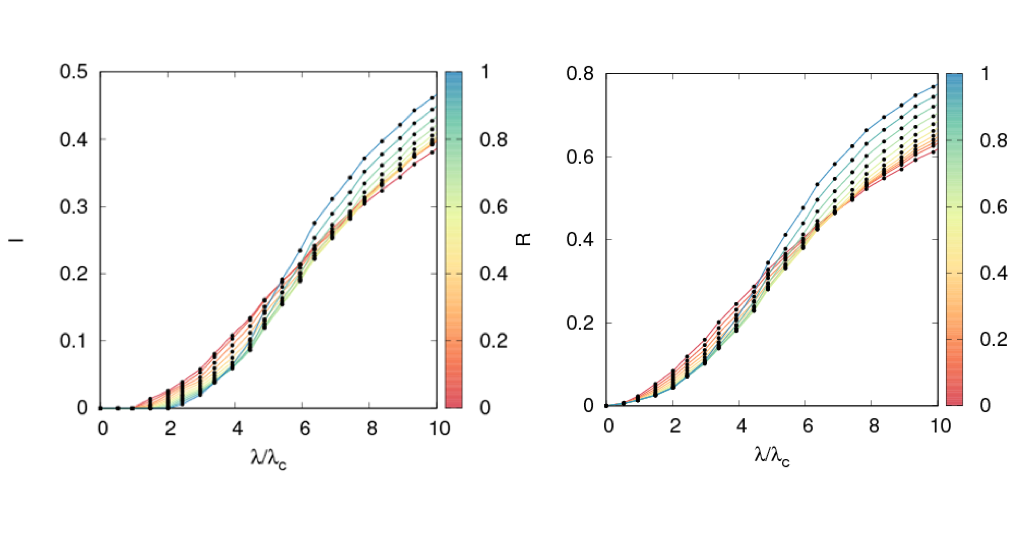}
\caption{Epidemic diagrams, $I(\lambda)$ (Left) and $R(\lambda)$ (Right) for SIR and SIS dynamics respectively in a real multiplex metapopulations. Solid lines denotes the predictions of our model about the incidence of a disease whereas black dots show the results obtained from averaging 20 realisation of numerical simulations. The colour code denotes the value of the mobility of the agents $p$. The recovery rate is set to $\mu=0.2$. }
\label{Fig8}
\end{figure}

\begin{figure*}[t!]
\begin{center}
\includegraphics[width=2.00\columnwidth]{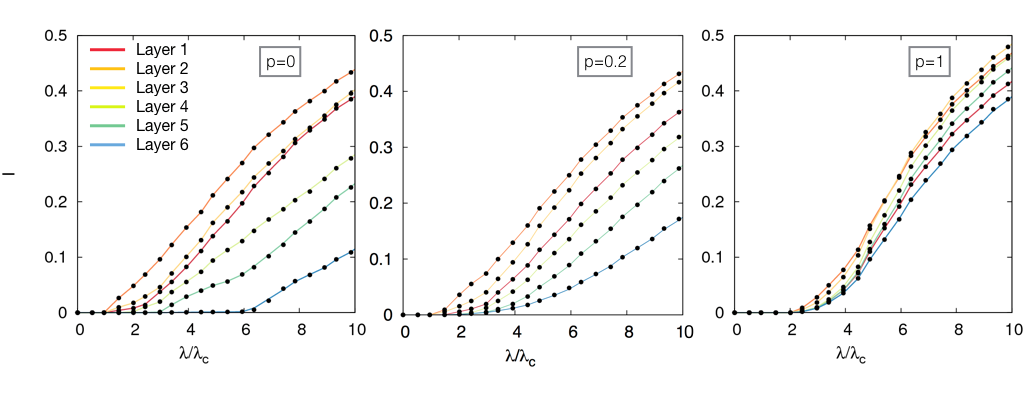}
\caption{\label{fig:Socio1} Impact of an SIS disease, $I(\lambda)$, on each of the layers of a real multiplex metapopulation. Note that the value of $\lambda$ has been re-scaled by the critical value $\lambda_c$ at $p=0$. The mobility of the agents has been set from left to the right to $p=0$, $p=0.2$ and $p=1$. Solid lines correspond to theoretical predictions obtained by iterating Eqs. \ref{eqrho}-\ref{travels}, whereas black dots are the result from averaging 20 realisations of numerical simulations. The recovery rate of both dynamics has been set to $\mu=0.2$. 
}
\label{Fig9}
\end{center}
\end{figure*}
To shed more light into the validity of the Markovian equations and, as a byproduct, to illustrate one relevant scenario where multiplex metapopulations capture contagion processes in real setups, we now study the SIS and SIR dynamics in an urban system (Medell\'{\i}n) where $6$ different socio-economic classes coexist. Specifically, these social classes range from 1, which gather those inhabitants with the lowest incomes, to class 6, corresponding to the wealthiest individuals. The separation into $6$ socioeconomic classes in Colombia \cite{colombia} and, in particular, in large cities such as Medell\'{\i}n (the second largest city in Colombia with around $5\cdot 10^{6}$ inhabitants) leads to a different demographic distribution across towns and, equally important, to different mobility patterns, as previously presented in \cite{socio1,socio2}.  

To study the evolution of diseases while preserving the information related to the existence of different socio-economic classes, we make use of the former formalism by constructing, from the data presented in \cite{source,data}, a multiplex network of $6$ layers. Each layer accounts for the mobility patterns as well as the distribution of the population which belong to each of the socioeconomic classes. Thus, the result is a multiplex network of $413\times 6$ nodes, where $413$ corresponds to the number of neighborhoods in which Medellin was divided for this study.

As in the case of synthetic networks, we first check the accuracy of our equations in predicting the total epidemic incidence. In Fig.~\ref{Fig8} we plot the epidemic diagrams corresponding to different values of the degree of mobility, $p$, for both SIS and SIR diseases. These diagrams are obtained via Monte Carlo simulations (points) and by solving the Markovian equations (lines), showing an  excellent agreement. Interestingly, we can also notice that, in Medell\'{\i}n, the agents mobility has a detrimental effect on the onset of epidemics, since the epidemic threshold increases with $p$. This counter-intuitive behaviour, which we have already reported for monolayer configurations \cite{natphys}, emerges from the homogenization of the demographic distribution while increasing the mobility.

The particular architecture of the Medellin multiplex allows us to assess the effect of the mobility on the impact of diseases within each social class. For this purpose, we represent in Fig.~\ref{Fig9}, for several values of the mobility $p$, the epidemic diagrams for a SIS disease, $I(\lambda)$. There we can find again a fair agreement between theory and simulations, what reveals that the multiplex Markovian equations allow to capture the incidence of epidemics at the layers' level. Remarkably, we observe that the critical point of the full multiplex moves towards high values (detrimental effect on the epidemic spreading \cite{natphys}) as mobility increases, but this effect is not reported on the individual layers by themselves. 

Remarkably, some features about the underlying multiplex network can be inferred from these graphs. For instance, the results corresponding to the static case ($p=0$) unveil the demographic distribution of the layers. On one hand, in Fig.~\ref{Fig9}.a, we can observe that agents from socioeconomic classes 1, 2 and 3 occupy the most populated nodes, since the epidemic onset associated to these layers is the smallest one. On the other hand, it becomes clear that individuals from social class 6 reside practically isolated from the rest of the classes, occupying sparsely populated neighbourhoods. Besides, from Fig.~\ref{Fig9}.b-c, we can also notice that mobility promotes the social mixing, since by increasing $p$ we obtain a more homogeneous impact of the disease across the layers.

To get more insight about the interaction among the different layers and to further validate our formalism, we now address the spatio-temporal propagation of diseases whose initial seed is localised inside one of the layers. For this purpose, we have fixed the parameters of our model ($p$, $\lambda$, $\mu$) and represented in Fig.~\ref{Fig10} the time evolution of the number of infected agents according to a SIR disease for each socioeconomic class when the seed is localised in classes 1 (Fig.~\ref{Fig10}.a) and 5 (Fig.~\ref{Fig10}.b). The solution of the Markovian equations captures the non-trivial interaction patterns between the different classes. In particular, it can be noticed that contagion processes take place mainly among close classes (in terms of incomes) since they show a cascade-like structure: $1\ \rightarrow\ 2,\; 3\  \rightarrow 4$ in Fig.~\ref{Fig10}.a. and $5\ \rightarrow \ 4 \ \rightarrow 3 \ \rightarrow 4,\; 2 \ \rightarrow \ 1$ in Fig.~\ref{Fig10}.b. Finally, the nontrivial nature of the time evolution of infections is captured by the existence of a feedback phenomenon when looking to the sequence of local outbreaks for classes $2$, $3$, and $4$. The observed correlations between layers' outbreaks reveals the closeness between the individuals in these middle class layers.

\begin{figure}[b!]
\begin{center}
\includegraphics[width=0.99\columnwidth]{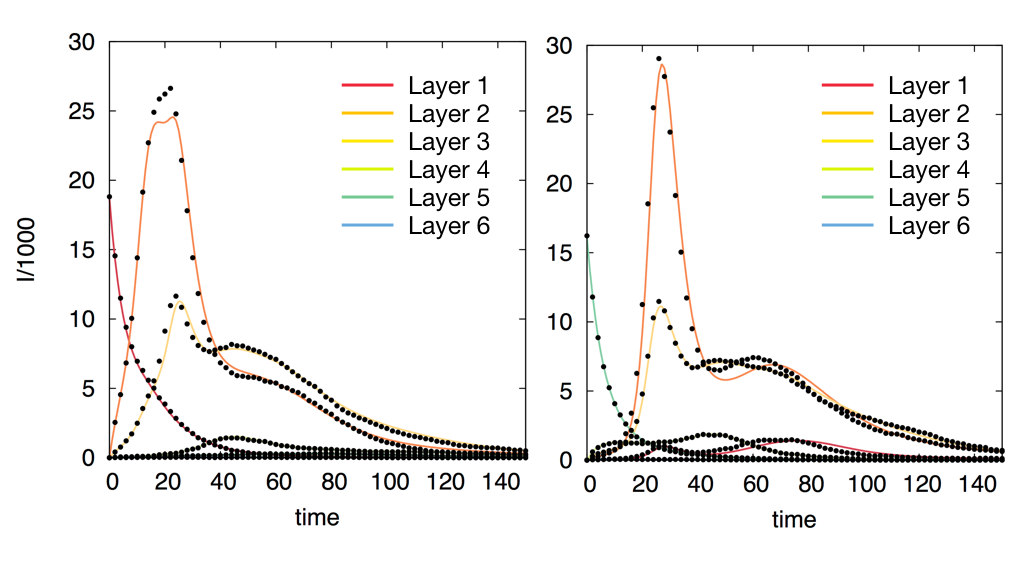}
\caption{\label{fig:Socio2} Temporal evolution of a SIR disease whose seed is initially localised inside layer 1 (Left) and layer 5 (Right). Solid lines correspond to theoretical predictions according to Eqs.(\ref{eqrho}-\ref{travels}) whereas black dots are the output of Monte Carlo simulations. The mobility of the agents $p$, the contagion rate $\lambda$ and the recovery rate $\mu$ have been set to ($p,\lambda,\mu$)= (0.05,4$\lambda_c$(p=0),0.2).  
}
\label{Fig10}
\end{center}
\end{figure}

}
\begin{figure*}[t!]
\begin{center}
\includegraphics[width=1.85\columnwidth]{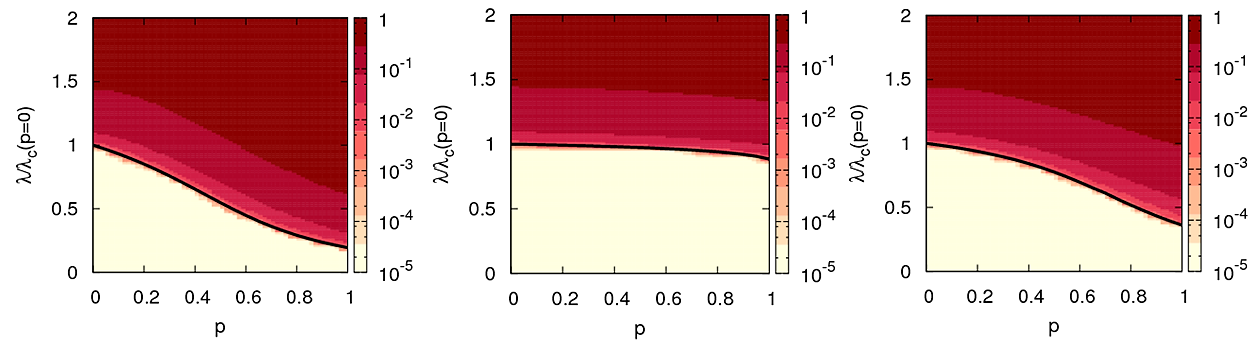}
\caption{\label{fig:umbralmultiplex} Epidemic diagrams, $I(\lambda,p)$ for SIS dynamics over those multiplex metapopulations used in Fig.\ref{fig:markovmultiplex}. The color-code (see color bar) denotes the fraction of infected individuals in the steady state as obtained from Monte Carlo simulations. The solid curves indicate the function $\lambda_c(p)$ as dictated from Eq. (\ref{multiplexthreshold}) for the multiplex metapopulations. The recovery rate of the SIS dynamics has been set to $\mu=0.2$. From left to the right we have used SF-SF, ER-ER and ER-SF architectures to simulate the evolution of a disease. 
}
\end{center}
\end{figure*}

\section{Deduction of the epidemic threshold}

The fair agreement between agent-based simulations and the solution of the Markovian equations allow us to make use of them in order to derive the analytical expression of the epidemic threshold. To get a first insight about the behavior of the epidemic threshold and its relation with the topology of the layers composing the multiplex network, we again consider that all the agents, regardless of their layers, interact in the same way so that $\lambda^{\alpha\beta}=\lambda \ \forall(\alpha,\beta)$. For the sake of simplicity, we also focus on the SIS case (similar results are obtained for the SIR model). In this case, see Eq. (\ref{eqrho}), the stationary solution for the fraction of infected agents of type $\alpha$ associated to patch $i$, $\rho_{i}^{\alpha\:\star}$, fulfills:
\begin{equation}
\mu\rho_i^{\alpha\:*}= (1-\rho_{i}^{\alpha\:*})\left((1-p)P^{\alpha\:*}_i + p\sum_{j=1}^{N}R^{\alpha}_{ij} P^{\alpha\:*}_{j}\right)\;.
\label{multiplexstat}
\end{equation}
As usual for calculating the threshold, we linearize the above expression by considering that the fraction of infected people in the stationary state is very small ($\rho_i^{\alpha:\star}= \epsilon_i^{\alpha} \:<<\:1 \: \forall\alpha \:\forall i$). This way, we can neglect second order terms in $\epsilon_i^{\alpha} $ in Eq (\ref{multiplexP}), so that $P^{\alpha\:*}_i$ is given by:
\begin{equation}
P^{\alpha\:*}_i=P^*_i=\sum_{\beta=1}^{L}\sum_{j=1}^{N}\lambda\epsilon^{\beta}_j n^{\beta}_{j\rightarrow i}\;.
\end{equation}
Introducing this expression into Eq.(\ref{multiplexstat}), the stationary state of the epidemics can be written as:
\begin{equation}
\mu\epsilon_i^{\alpha} =(1-p)\sum_{\beta=1}^{L}\sum_{j=1}^{N}\lambda\epsilon^{\beta}_j(t)n^{\beta}_{j\rightarrow i} + p\sum_{j=1}^{N}R^{\alpha}_{ij}\sum_{\beta=1}^{L}\sum_{k=1}^{N}\lambda\epsilon^{\beta}_k n^{\beta}_{k\rightarrow j}\;.  
\end{equation}
By using the value of $n^{\beta}_{j\rightarrow i}$ from Eq. (\ref{travels}) and keeping up to first order in $\epsilon_i^{\alpha}$ we finally obtain the expression:
\begin{widetext}
\begin{equation}
\frac{\mu}{\lambda}\epsilon_i^{\alpha} = \sum_{j=1}^N\sum_{\beta=1}^{L}\underbrace{\left[(1-p)^{2}\delta_{ij}n_{i}^{\beta} +p(1-p)n_{j}^{\beta}(R_{ji}^{\beta}+R_{ij}^{\alpha}) + p^2 n_{j}^{\beta}(\textbf{R}^{\alpha}\cdot\textbf{R}^{\beta\; T})_{ij}\right]}_{{\cal M}_{ij}^{\alpha\beta}}\epsilon_j^\beta \;.
\label{metamatrix}
\end{equation}
\end{widetext}
 
At this point, it becomes clear that Eq. (\ref{metamatrix}) defines an eigenvalue problem for the feasible solutions $\epsilon_{i}^{\alpha}$. Indeed, there are $N\cdot L$ feasible solutions of $\lambda$ corresponding to the eigenvalues of the $N\cdot L\times N\cdot L$ supra-matrix ${\cal M}$. However, since we are interested in the minimum value $\lambda_c$ for which Eq. (\ref{metamatrix}) is fulfilled, the epidemic threshold is thus associated to the largest eigenvalue of ${\cal M}$ as:
\begin{equation}
\lambda_c=\frac{\mu}{\Lambda_{max} ({\cal M})}\;.
\label{multiplexthreshold}
\end{equation}

Let us now describe the entries of the matrix ${\cal M}$, see Eq. (\ref{metamatrix}), since they allow us to quantify the microscopical interactions among agents across the multiplex metapopulations. In fact, the elements ${\cal M}_{ij}^{\alpha\beta}$ correspond, close to the epidemic threshold, to the probability that an agent of type $\alpha$ associated to patch $i$ contacts with another one of type $\beta$ from patch $j$. Specifically, each element contains three contributions accounting for the three potential sources of infections that a healthy agent can find: from agents associated to the same node inside this node [weighted by $(1-p)^2$], from agents from a different patch, either at one of the two patches they are associated to [weighed by $p(1-p)$], and from agents with whom she contacts inside a third place different from their associated nodes (weighted by $p^2$).


To round off this derivation, let us remark that the generality of the expression for the epidemic threshold of multiplex metapopulations allows us to recover, by setting $L=1$, the value of the epidemic threshold for diseases which propagates over single metapopulations. Indeed, for $L=1$, Eq.~(\ref{metamatrix}) turns into:
\begin{widetext}
\begin{equation}
\frac{\mu}{\lambda}\epsilon^*_i=\sum_{j=1}^{N}\underbrace{\left[ (1-p)^{2}\delta_{ij}n_{j}+p(1-p)n_{j}
\left({\bf R+R^T}\right)_{ij} 
+ 
p^{2}n_{j}\left({\bf R\cdot R^T}\right)_{ij}\right]}_{M_{ij}}\epsilon^{*}_{j}\;,
\label{matrixM}
\end{equation}
\end{widetext}
so that the epidemic threshold is given by:
\begin{equation}
\lambda_{c} = \frac{\mu}{\Lambda_{max} ({\bf M})}\;,
\label{threshold}
\end{equation}
where $M$ is now an $N\times N$ matrix. In the same fashion as supra-matrix ${\cal M}$, each term $M_{ij}$ of matrix ${\bf M}$ encodes the probability that an agent associated to patch $i$ contacts with another from patch $j$.

We have checked the validity of Eq. (\ref{multiplexthreshold})  by computing the largest eigenvalue of ${\cal M}$ for the three synthetic multiplexes under study in Fig. (\ref{fig:markovmultiplex}) for a range of values of $p\in[0,1]$. This way, through Eq. (\ref{multiplexthreshold}) we obtain a curve $\lambda_c(p)$, see Fig. \ref{fig:umbralmultiplex}, that reproduces the onset observed in Monte Carlo simulations.  The monotonous decreasing behavior of curve $\lambda_c(p)$ corroborates that mobility enhances the spread of the disease.

\section{Conclusions}

In this work, we have elaborated a theoretical formalism to analyze spreading processes in multiplex metapopulations characterized by recurrent mobility patterns. Our framework gets rid of the assumptions about the correlations between the node attributes and epidemic variables introduced in heterogeneous mean field formulations. This way, the formalism introduced here is general enough so to accommodate any origin-destination (weighted and directed) matrix containing different commuting patterns within a population and to cast the information about the local census of each patch.

First, we have introduced the Markovian evolution equations for the monoplex (single layer multiplex) case under the SIR and SIS dynamics. We have tested their validity by solving these equations and comparing their solution with the results obtained from Monte Carlo simulations in synthetic ER and SF networks. The agreement obtained is remarkable both at the macroscopic and the microscopic level. In particular, by solving the Markovian equations we can reproduce the spatio-temporal epidemic patterns capturing the onset of epidemics at the local level of patches. The second step has been to generalize the former formalism to address metapopulations composed of several types of agents whose mobility patterns are different. To this aim, we have made use of the multiplex formalism, thus constructing a multiplex metapopulation. We have again checked the validity of the Markovian formalism showing a great accuracy for both macroscopic and microscopic indicators.

The validity of the Markovian equations has allowed us to derive analytical expressions for the global epidemic threshold of multiplex metapopulations. Again, the analytical prediction is in complete agreement with numerical simulations. Interestingly, the onset is related to the maximum eigenvalue of a supra-matrix ${\cal M}$ in which the different mobility patterns, local census and the degree of mobility interplay. Remarkably, the structure of these supra-matrix ${\cal M}$ captures three basic contagion processes for a healthy individual. 

On more general grounds, dynamical processes on multiplexes have been a research focus in the recent years \cite{diffusion,games,patterns,synchronization} and, in particular, their application to epidemics \cite{epid1}. As usual in the multiplex literature, the scenario considered is that of coupled contact networks, so that a node is an individual that interact in different ways  ({\it i.e.} through different interaction layers) with the rest of the nodes. Under this setting, different problems such as the diffusion of a disease through different contagion channels \cite{cozzo,buono,arruda}, the cooperative spreading of different diseases \cite{sanz,sahneh,nahid,fakhteh} or the coevolution of different contagion processes \cite{granell1,granell2} have been addressed. Here, at variance, the two interaction levels (epidemics and mobility) of the metapopulation yield interesting results related to the interplay of the architecture of layers. 
We have shown the applicability of the formalism to a real case study (city of Medellin) where we have gathered data of the mobility patterns for different socio-economical classes (layers). The results present an interesting behaviour of epidemic detriment \cite{natphys} for the full multiplex structure while there is no epidemic detriment for all individual layers.

In a nutshell, the formalism introduced here provides with a reliable and computationally time-saving platform to analyze the epidemic risk of systems displaying recurrent mobility patterns. This way, the formalism can be used to readily identify those critical areas that spur the unfolding of diseases. In addition, the possibility of handling analytical equations can be further exploited beyond the derivation of the epidemic threshold and combined together with control techniques to test in an efficient way different contention policies. We expect as well that our Markovian formalism can be further extended in the future to accommodate more sophisticated commuting patterns and more refined epidemic models, thus approaching more to real epidemic scenarios.  

\section{Acknowledgements}

We acknowledge useful discussions and suggestion by S. Meloni. Financial support came from MINECO (through projects FIS2015-71582-C2 and FIS2014-55867-P), from the Departamento de Industria e Innovaci\'on del Gobierno de Arag\'on y Fondo Social Europeo (FENOL group E-19).
AA acknowledges ICREA Academia and the James S.\ McDonnell Foundation.


\begin{thebibliography}{}


\bibitem{gleam1} D. Balcan, B. Gonçalves, H. Hu, J. J. Ramasco, V. Colizza and A.Vespignani. Modeling the spatial spread of infectious diseases: The GLobal Epidemic and Mobility computational model. {\it Journal of Computational Science} {\bf 1}, 132--145 (2010).

\bibitem{gleam2} M. Tizzoni {\it et al.} Real-time numerical forecast of global epidemic spreading: case study of 2009 A/H1N1pdm. {\it BMC Medicine} {\bf 10} 165 (2012). 

\bibitem{hanski} I. Hanski.  Metapopulation dynamics. {\it Nature} {\bf 396}, 41--49 (1998).

\bibitem{metapop1} I. Hanski, and M.E. Gilpin. {\it Metapopulation Biology: Ecology, Genetics, and Evolution}. (Academic Press, 1997).

\bibitem{metapop2} D. Tilman and P. Kareiva. {\it Spatial Ecology}. (Princeton University Press, 1997).

\bibitem{metapop3} J. Bascompte,  and  R.V. Sol\'e. {\it Modeling Spatiotemporal Dynamics in Ecology}. (Springer, 1998).

\bibitem{metapop4} I. Hanski, and O.E. Gaggiotti. {\it Ecology, Genetics, and Evolution of Metapopulations}.(Elsevier and Academic Press, 2004).

\bibitem{metapop5} M.J. Keeling, and P. Rohani. {\it Modeling Infectious Diseases in Humans and Animals}. (Princeton University Press, 2008).

\bibitem{pion1} L. Sattenspiel, and K. Dietz. A structured epidemic model incorporating geographic mobility among regions. {\it Mathematical Biosciences} {\bf 128}, 71--91 (1995).

\bibitem{pion2} B. Grenfell, and J. Harwood. (Meta)population dynamics of infectious diseases. {\it Trends in Ecology \& Evolution} {\bf 12}, 395--399. (1997).

\bibitem{airline} R. Guimer\'a, S. Mossa, A. Turtschi, and L. A. N. Amaral The worldwide air transportation network: Anomalous centrality, community structure, and cities' global roles. {\it Proc. Nat. Acad. Sci (USA)}  {\bf 102} 7794--7799 (2005).

\bibitem{review} R. Pastor-Satorras, C. Castellano, P. Van Mieghem, and A. Vespignani. Epidemic processes in complex networks. {\it Rev. Mod. Phys.} {\bf 87}, 925--979 (2015).


\bibitem{colizza1} V. Colizza, R. Pastor-Satorras, and A. Vespignani. Reaction--diffusion processes and metapopulation models in heterogeneous networks. {\it Nature Phys.} {\bf 3}, 276--282 (2007).

\bibitem{colizza2} V. Colizza, and A. Vespignani. Invasion threshold in heterogeneous metapopulation networks. {\it Phys. Rev. Lett.} {\bf 99}, 148701 (2007).

\bibitem{colizza3} V. Colizza, and A. Vespignani. Epidemic modeling in metapopulation systems with heterogeneous coupling pattern: Theory and simulations. Journal of theoretical biology {\bf 251}, 450--467 (2008).

\bibitem{colizza4} D. Balcan, V. Colizza, B. Gon�alves, H. Hao, J.J. Ramasco, and A. Vespignani. Multiscale mobility networks and the spatial spreading of infectious diseases- {\it Proc. Nat. Acad. Sci (USA)} {\bf 106}, 21484--21489 (2009).

\bibitem{eubank} S. Eubank, {\it et al}. Modelling disease outbreaks in realistic urban social networks. {\it Nature} {\bf 429}, 180--184 (2004).

\bibitem{hmf} R. Pastor-Satorras, and A. Vespignani. Epidemic Spreading in Scale-free networks. {\it Phys. Rev. Lett.} {\bf 86}, 3200 (2001).

\bibitem{influenzanet} C. Guerrisi {\it et al.} Participatory Syndromic Surveillance of Influenza in Europe. {\it Journal of Infectious Diseases} {\bf 214} S386-S392 (2016). 

\bibitem{ball} F. Ball {\it et al}. Seven challenges for metapopulation models of epidemics, including households models. {\it Epidemics} {\bf 10}, 63--67 (2015).

\bibitem{commutes} D. Balcan, and A. Vespignani. Phase transitions in contagion processes mediated by recurrent mobility patterns. {\it Nature Phys.} {\bf 7}, 581--586
(2011).

\bibitem{prx} V. Belik, T. Geisel and D. Brockmann. Natural Human Mobility Patterns and Spatial Spread of Infectious Diseases. {\it Phys. Rev.~X} {\bf 1}, 1, 011001 (2011).

\bibitem{jtb} D Balcan, and  A Vespignani. Invasion threshold in structured populations with recurrent mobility patterns. {\it J. Theor. Biol.} {\bf 293}, 87--100 (2011).

\bibitem{epjb} V. Belik, T. Geisel and D. Brockmann. Recurrent host mobility in spatial epidemics: beyond reaction-diffusion. {\it Eur. Phys. J. B} {\bf 84}, 579--587 (2011).

\bibitem{kivela2014} M. Kivel\"a, A. Arenas, M. Barthelemy, J.P. Gleeson, Y. Moreno, and M.A. Porter. Multilayer Networks. {\it J. Complex Networks} {\bf 2}, 203-271 (2014)

\bibitem{boccaletti2014} S. Boccaletti, G. Bianconi, R. Criado, C.I. del Genio, J. G\'omez-Garde\~nes, M. Romance, I. Sendi\~na-Nadal,
Z. Wangk, and M. Zanin. The structure and dynamics of multilayer networks. {\it Phys. Rep.} {\bf 544}, 1--122 (2014)

\bibitem{dedomenico2013} M. De Domenico,  A. Sol\'e-Ribalta, E. Cozzo, M. Kivel\"a, Y. Moreno, M.A. Porter, S. G\'omez and A. Arenas. Mathematical formulation of multi-layer networks. {\it Phys. Rev. X} {\bf 3}, 041022 (2013).

\bibitem{battiston} F. Battiston, V. Nicosia, and V. Latora. Structural measures for multiplex networks. {\it Phys Rev E} {\bf 89}, 032804 (2014).

\bibitem{radicchi} F. Radicchi, and A. Arenas. Abrupt transition in the structural formation of interconnected networks. {\it Nature Phys.} {\bf 9}, 717--720 (2013).

\bibitem{natphys} J. G\'omez-Garde\~nes, D. Soriano-Pa\~nos, and A. Arenas. Critical regimes driven by recurrent mobility patterns of reaction-diffusion processes in networks. {\it Nature Phys.} (doi: 10.1038/s41567-017-0022-7) (2018).

\bibitem{scarpino} S. V. Scarpino. Don't close the gates. {\it Nature Phys.} (doi: 10.1038/s41567-017-0028-1) (2018).

\bibitem{EPL} S. G\'omez, A. Arenas, J. Borge-Holthoefer, S. Meloni, and Y. Moreno. Discrete-time Markov chain approach to contact based diseases spreading in complex networks. {\it EPL} {\bf 89}, 38009 (2010).

\bibitem{Guerra} B. Guerra, and J. G\'omez-Garde\~nes, Annealed and mean-field formulations of disease dynamics on static and adaptive networks. {\it Phys. Rev. E} {\bf 82}, 035101 (2010).

\bibitem{PRE} S. G\'omez, J. G\'omez-Garde\~nes, Y. Moreno and A. Arenas. Nonperturbative heterogeneous mean-field approach to epidemic spreading in complex networks. {\it Phys. Rev. E} {\bf 84} 036105 (2011).

\bibitem{cozzo} E. Cozzo, R.A. Ba\~nos, S. Meloni, and Y. Moreno. Contact-based social contagion in multiplex networks. {\it Phys. Rev E} {\bf 88}, 050801 (2013).

\bibitem{arruda} G.F. de Arruda, E. Cozzo, T.P. Peixoto, F.A. Rodrigues, and Y. Moreno. Disease Localization in Multilayer Networks. {\it Phys. Rev. X} {\bf 7}, 011014 (2017).

\bibitem{colombia} C. Medina, L. Morales, R. Bernal, and M. Torero. Stratification and Public Utility Services
in Colombia: Subsidies to Households or Distortion of Housing Prices? {\it Economia} {\bf 7}, 41--99 (2007).

\bibitem{socio1} L. Lotero, R.G. Hurtado, L.M. Flor\'{\i}a, and J. G\'omez-Garde\~nes. Rich do not rise early: Spatio-temporal patterns in the Mobility Networks of different Socio-economic classes. {\it Royal Society Open Science} {\bf 3}, 150654 (2016).

\bibitem{socio2}  L. Lotero, A. Cardillo, R.G. Hurtado, and J. G\'omez-Garde\~nes. Several Multiplexes in the same city: The role of Wealth differences in urban mobility, in {\it Interconnected Networks} (Springer, 2016), pp. 149--164.

\bibitem{source} Universidad Nacional de Colombia and AREA Metropolitana del Valle de Aburr\'a 2006. Encuesta origen destino de viajes 2005 del Valle de Aburrá, estudios de tr\'ansito complementarios y validaci\'on. (Technical Report).

\bibitem{data} L. Lotero, R.G. Hurtado, L.M. Flor\'{\i}a, and J. G\'omez-Garde\~nes. Rich do not rise early: Spatio-temporal patterns in the Mobility Networks of different Socio-economic classes. Dryad Digital Repository. (http://dx.doi.org/10.5061/dryad.hj1t4).

\bibitem{diffusion} S. G\'omez, {\it et al}. Diffusion Dynamics on Multiplex Networks. {\it Phys. Rev. Lett.} {\bf 110}, 028701 (2013).

\bibitem{games}  J. G\'omez-Garde\~nes, I. Reinares, A. Arenas, LM Flor\'{\i}a. Evolution of cooperation in multiplex networks. {\it Sci. Rep.} {\bf  2}, 620 (2012).

\bibitem{patterns} N.E. Kouvaris, Shigefumi Hata, and Albert D\'{\i}az-Guilera. Pattern formation in multiplex networks. {\it Sci. Rep.} {\bf 5}, 10840 (2015).

\bibitem{synchronization} C.I. Del Genio, J. G\'omez-Garde\~nes, I. Bonamassa, and S. Boccaletti. Synchronization in networks with multiple interaction layers. {\it Science Adv.} {\bf 2}, e1601679 (2016).
 
\bibitem{epid1} A. Saumell-Mendiola, M.A. Serrano, and M. Bogu\~n\'a. Epidemic spreading on interconnected networks. {\it Phys. Rev. E} {\bf 86}, 026106 (2012).

\bibitem{buono} C. Buono, L.G. Alvarez-Zuzek, P.A. Macri, and L.A. Braunstein. Epidemics in partially overlapped multiplex networks. {\it PloS One} {\bf 9}, e92200 (2014).

\bibitem{sanz} J. Sanz, Ch.-Y. Xia, S. Meloni, and Y. Moreno. Dynamics of Interacting Diseases. {\it Phys. Rev. X} {\bf 4}, 041005 (2014).

\bibitem{sahneh} F.D. Sahneh, and C. Scoglio. Competitive epidemic spreading over arbitrary multilayer networks. {\it Phys. Rev. E} {\bf 89}, 062817 (2014).

\bibitem{fakhteh} W. Cai, L. Chen, F. Ghanbarnejad, and P. Grassberger. Avalanche outbreaks emerging in cooperative contagions. {\it Nature Phys.} {\bf 11}, 936 (2015).

\bibitem{nahid} N. Azimi-Tafreshi. Cooperative epidemics on multiplex networks. {\it Phys. Rev. E} {\bf 93}, 042303 (2016).

\bibitem{granell1} C. Granell, S. G\'omez, and A. Arenas. Dynamical Interplay between Awareness and Epidemic Spreading in Multiplex Networks. {\it Phys. Rev. Lett.} {\bf 111}, 128701 (2013).

\bibitem{granell2} C. Granell, S. G\'omez, and A. Arenas. Competing spreading processes on multiplex networks: Awareness and epidemics. {\it Phys. Rev. E} {\bf 90}, 012808 (2014).

\end{thebibliography}
\end{document}